\documentclass[fleqn,usenatbib]{mnras}

\usepackage{newtxtext,newtxmath}
\usepackage{pdflscape}
\usepackage{soul} 
\usepackage{cancel} 

\usepackage[T1]{fontenc}

\DeclareRobustCommand{\VAN}[3]{#2}
\let\VANthebibliography\thebibliography
\def\thebibliography{\DeclareRobustCommand{\VAN}[3]{##3}\VANthebibliography}

\usepackage{graphicx}	
\usepackage{amsmath}	
\usepackage{multirow}
\usepackage{float}
\usepackage{xcolor}

\newcommand{\vtheta}{\boldsymbol{\theta}}

\newcommand{\aref}[1]{\hyperref[#1]{Appendix~\ref{#1}}}

\newcommand{\nclus}{1178}

\title[Cluster Demographics in NGC 628]{Cluster Population Demographics in NGC 628 Derived from Stochastic Population Synthesis Models}

\author[Tang et al.]{Jianling Tang,$^{1,2}$\thanks{E-mail: janett.jianling@gmail.com, u6735916@anu.edu.au}
Kathryn Grasha,$^{1,2,3}$\thanks{ARC DECRA Fellow}
Mark R. Krumholz$^{1,2}$
\\
$^{1}$Research School of Astronomy and Astrophysics, Australian National University, Canberra, ACT 2611, Australia\\
$^{2}$ARC Centre of Excellence for All Sky Astrophysics in 3 Dimensions (ASTRO 3D), Australia\\
$^{3}$Visiting Fellow, Harvard-Smithsonian Center for Astrophysics, 60 Garden Street, Cambridge, MA 02138, USA\\
}

\date{Accepted 2024 Jul 22. Received 2024 Jul 22; in original form 2023 Jan 15}

\pubyear{2024}

\begin{document}
\label{firstpage}
\pagerange{\pageref{firstpage}--\pageref{lastpage}}
\maketitle

\begin{abstract}
The physical properties of star cluster populations offer valuable insights into their birth, evolution, and disruption. However, individual stars in clusters beyond the nearest neighbours of the Milky Way are unresolved, forcing analyses of star cluster demographics to rely on integrated light, a process fraught with uncertainty. Here we infer the demographics of the cluster population in the benchmark galaxy NGC 628 using data from the Legacy Extra-galactic UV Survey (LEGUS) coupled to a novel Bayesian forward-modelling technique. Our method analyzes all \nclus\ clusters in the LEGUS catalogue, $\sim4$ times more than prior studies severely affected by completeness cuts. Our results indicate that the cluster mass function is either significantly steeper than the commonly-observed slope of $-2$ or is truncated at $\approx 10^{4.5}$ M$_\odot$; the latter possibility is consistent with proposed relations between truncation mass and star formation surface density. We find that cluster disruption is relatively mild for the first $\approx 200$ Myr of cluster evolution; no evidence for mass-dependent disruption is found. We find suggestive but not incontrovertible evidence that inner galaxy clusters may be more prone to disruption and outer galaxy clusters have a more truncated mass function, but confirming or refuting these findings will require larger samples from future observations of outer galaxy clusters. Finally, we find that current stellar track and atmosphere models, along with common forms for cluster mass and age distributions, cannot fully capture all features in the multidimensional photometric distribution of star clusters. While our forward-modelling approach outperforms earlier backward-modeling approaches, some systematic differences persist between observed and modelled photometric distributions.
\end{abstract}

\begin{keywords}
galaxies: star clusters: general --  techniques: photometric  -- software: data analysis -- methods: statistical -- galaxies: individual (NGC 628) -- galaxies: star formation
\end{keywords}



\section{Introduction}
The distribution of star clusters' properties (mass, age, physical size) and the variation of this distribution with galactic environment provides crucial clues to the physics of star formation and star cluster evolution. Moreover, because of the processes responsible for setting this distribution may depend on galactic environment, cluster demographics can also trace the history of galaxy assembly and evolution, acting as a fossil record of the environments that existed when clusters formed
\citep[e.g.,][]{2015Galactic_var, 2021Menon}. Given the importance of cluster demographics, it is not surprising that there have been many attempts to measure them, and that cluster demographics figure prominently in the science cases for a number of large surveys of nearby galaxies, such as PHAT \citep{2012PHAT}, PHANGS-HST \citep{2022PHANGS}, and LEGUS \citep{2015LEGUS}; see \citet{2019ARAA} for a detailed review.


Different aspects of cluster demographics probe different physics.
By studying the shape of the cluster mass function (CMF), and in particular the initial CMF (ICMF) that applies to the youngest clusters, we can constrain star formation theories and provide an observational check on simulations. 
Studies to date show that over much of its range the ICMF is well-described by a powerlaw $dN/dM\propto M^{\alpha_M}$ with a slope $\alpha_M \approx -2$ across a wide range of galaxy properties, corresponding to equal mass per logarithmic bin and suggesting a scale-free formation process \citep[e.g.,][]{2012Fall_MID,2014Chandar_MID}.
However, the shape of the ICMF at its high-mass end ($\gtrapprox 10^4 M_{\odot}$) remains uncertain. Some authors report that a truncated distribution such as a Schechter function describes the data better than the pure powerlaw \citep[e.g.,][]{2012Bastian_Schetcher, LEGUS2017, 2017PHAT}, while others question the statistical significance of claimed detections and instead suggest that the dearth of massive clusters is simply a size-of-sample effect \citep[e.g.,][]{Larsen09, 2020Mok}. If there is a real truncation in the mass function, its location must depend somehow on the galactic environment, since rapidly star-forming galaxies with large cluster populations often harbour clusters with masses larger than the reported truncation masses in more modestly star-forming galaxies \citep[e.g.,][]{2015MID, 2017Linden_inclusive}. The possible absence of a high-mass break in the ICMF, as well as its value and variations with the galactic environment, can provide important clues about the process of star formation.


As clusters evolve and disperse from their birthplaces, they experience mass loss. Various theories govern how they disrupt, ranging from the ``infant mortality'' stage of rapid gas removal after star formation \citep{1980Hills,2007gas_expulsion,2020gas_expulsion} to long-timescale processes such as two-body relaxation \citep{2001twobody,2010two_body,2012two_body} to processes such as external tidal shocking that operate on intermediate timescales \citep{2005Lamers_tidal,2005Bastian_tidal,2006GMC_encounter,2016tidal,2017GMC_encounter,2019tidal,2019Webb_tidal}. 
To place observational constraints on these theories, we study 
the cluster age function (CAF). 
As with the ICMF, observations suggest that the CAF is reasonably well-described by a powerlaw form $dN/dT\propto T^{\alpha_T}$, where $T$ is cluster age. However, the value of $\alpha_T$ remains highly controversial, partly due to disagreement among observational groups about what constitutes a cluster \citep{2019ARAA}. 
Despite this uncertainty, the CAF encodes the timescales of diverse cluster disruptive processes. 
One question of particular interest is the form of
the joint mass-age distribution, which gives a complete description of cluster formation and disruption.
If the mass and age distributions are separable, i.e.~if $d^2 N/dM\,dT \propto (dN/dM)(dN/dT)$, this implies
mass-independent disruption (MID) of clusters \citep{2005MID,2009MID,2015MID}, while if they are not this implies mass-dependent disruption (MDD; \citealt{2005MDD,2007MDD,2021MDD}).
Knowing whether MID or MDD holds in a particular galaxy or sub-galactic region would in turn place strong constraints on the mechanisms by which clusters disrupt.

Part of the reason that observations have not yet settled debates about the existence and value of a cutoff in the ICMF, the slope of the CAF, and whether cluster disruption is mass-dependent or -independent, is that determining the properties of clusters from observations is not trivial. To sample a wide range of environments,
studies of cluster populations must work with integrated light rather than resolved stellar populations, since in the crowded environments of star clusters it is generally only possible to resolve individual stellar sources for a handful of the most nearby galaxies.
The traditional approach to extracting cluster demographics from this type of data is to convert the integrated light measurements for each cluster to physical properties such as mass and age by comparing the observed photometry to a grid of simple stellar population (SSP) models and generating a set of best-fitting parameters. One then bins the clusters by mass and age to obtain mass and age distributions. 

However, this approach encounters several difficulties. First, the binning process usually entails the loss of useful information, and as a result parameters determined by fitting data to histograms are heavily biased \citep{2009Binning,2017Binning}. Second, the conventional approach of using $\chi^2$ fitting to convert photometry to masses and ages implicitly assumes that the uncertainties on cluster mass and age can be approximated as Gaussian. This is often a poor assumption, because the mapping between physical properties and photometry is non-linear and non-monotonic, particularly once one adds the additional complication of dust extinction. Consequently, a particular set of photometric measurements may plausibly fit two (or more) widely-separated loci in physical parameter space, yielding posterior mass and age distributions with complex non-symmetrical and multi-modal shapes \citep{2014deMeu_MTshape,2014MT_shape,2015Powerlaw,2019MT_shape}. 
The problem is exacerbated for low-mass clusters, where stochastic sampling of the IMF leads to a wide range of possible photometry even for clusters of fixed mass and age \citep{2009Piskunov}. 

However, the most severe issue for measuring cluster demographics is completeness. The problem is that star clusters become fainter as one moves to both lower mass and older age, so a magnitude limit corresponds to a complex shape in the parameter space of age and mass, meaning that both the mass and age distributions are subject to large potential biases. The conventional way of handling this is to discard clusters outside of a limited range of both mass and age over which completeness is expected to be $\gtrsim 90 \% $, e.g., to retain only clusters with estimated masses $5000$ ~M$_\odot$ and estimated ages $<200$~Myr in LEGUS. Such a drastic truncation of the sample avoids bias, but at the cost of a large loss of statistical power at both lower masses and old ages. The former is particularly concerning, since the steep mass function means that low-mass clusters form the majority of the available sample; the sample truncation required to avoid bias from survey magnitude limits therefore often involves discarding the majority of the data.

The main objective of this paper is to present a complete analysis pipeline and preliminary results for cluster demographics in the benchmark galaxy NGC 628 using the catalogue of clusters measured by the \emph{Hubble Space Telescope} Treasury Program Legacy Extragalactic Ultraviolet Survey \citep[LEGUS;][]{2015LEGUS} coupled to the novel Bayesian forward modelling method proposed by \citet{SLUG2019} that overcomes the problems identified above. Specifically, this method naturally copes with complex and non-deterministic mappings between photometric measurements and physical properties, and it enables us to use a total of \nclus\ clusters catalogued by LEGUS with non-zero completeness values, as compared to earlier modelling where severe completeness cuts reduced the sample to $\approx 300$ \citep{LEGUS2017}. 
This paper is structured as follows: in \autoref{cha:data}, we introduce our target galaxy, cluster catalogue, and analysis of completeness. In \autoref{cha:methods}, we summarise our analysis method. We present our results for cluster demographics in \autoref{cha:results}, and discuss their implications in \autoref{cha:discussions}. We summarise the findings and discuss future prospects in \autoref{cha:conclusions}.

\section{Data Description}
\label{cha:data}

Here we summarise the properties of our target galaxy (\autoref{ssec:ngc628}), the star cluster catalogue we use as the basis for our study (\autoref{ssec:classification}), and our treatment of survey completeness (\autoref{ssec:completeness}).

\subsection{NGC 628}
\label{ssec:ngc628}

\begin{figure}
  \centering
  \includegraphics[width=\columnwidth]{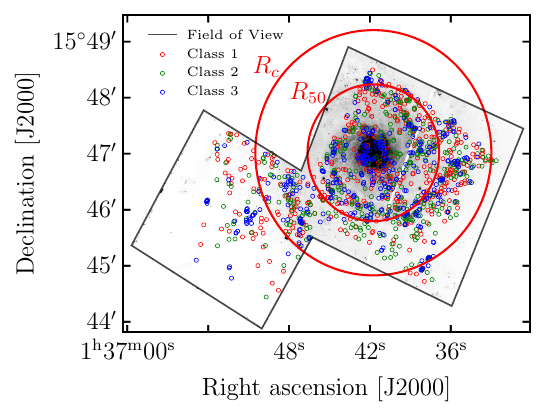}
 \caption{V-band (F555W) mosaic image of the centre and east fields of NGC 628 from LEGUS-HST. Symbols show the location of the star clusters, coloured by the cluster classifications: class 1 (red circle), class 2 (green circle), and class 3 (blue circle). The black boxes outline the observational field-of-view. For a detailed description of cluster classification method, see \autoref{ssec:classification}. The two large red circles represent the co-rotation radius $R_c$ (\autoref{ssec:cordiv}), and the median galactocentric radius of the LEGUS star cluster catalogue $R_{50}$ (\autoref{ssec:eqndiv}).}
  \label{fig:NGC628_cl}
\end{figure}

Our target NGC 628, also known as Messier 74, is a well-studied grand-design spiral galaxy, located at 9.9 Mpc \citep{2010Olivares}. We choose NGC 628 for this study as its large size (apparent radius of 5.2'; \citealt{2014Gusev}) and face-on orientation allow for a detailed study of the stellar cluster population. We use observations of NGC 628 taken from the LEGUS survey, a Cycle 21 HST Treasury programme that targeted 50 local galaxies ($\lesssim$15~Mpc) with the Hubble Space Telescope with broad band filter coverage from the UV to the near IR. All targets were imaged with either the Wide Field Camera 3 (WFC3) or the Advanced Camera for Surveys (ACS) in the NUV (WFC3 F275W), U-band (F336W), B-band (ACS/F435W or WFC3/F438W), V-band (ACS or WFC3/F555W or F606W), and I-band (ACS or WFC3 F814W). For the remainder of this paper we follow the conventional Johnson passband naming convention UV, U, B, V and I without converting to the Johnson system.
As part of the LEGUS survey, NGC~628 is covered by two pointings; one at the galactic centre (NGC 628c) and the other at the eastern edge (NGC 628e). We show the LEGUS V-band image of NGC~628 in \autoref{fig:NGC628_cl}. Detailed descriptions of the survey and data reduction are provided in \citet{CalzettiLEGUS2015}.

\subsection{Cluster Catalogue}
\label{ssec:classification}

The photometric star cluster catalogue for NGC~628 that we use in this study comes from the LEGUS survey, and we refer readers to \citet{Grasha2015} and \citet{LEGUS2017} for a complete description of how it is constructed. Here, we provide a brief description of the parts of the process most pertinent to this study. LEGUS uses a two-step pipeline for cluster classification. First, an automated cluster extraction tool is used to extract potential cluster candidates with at least a 3$\sigma$ detection in a minimum of 5 contiguous pixels from the white light images. The cluster candidates are then further refined and required to have a V-band concentration index (CI; the difference in magnitude between a radius of 1 and 3 pixels) of $>1.4$ mag for the centre pointing and $>1.3$ mag for the east pointing. The CI cut serves to separate point-like sources (i.e., stars) from more extended sources (i.e., star clusters). The star cluster candidates are required to be detected in the V-band and a minimum of at least three of the other filters, with photometric error $\sigma_\lambda\leq 0.3$ mag in each band.

To create the final cluster catalog, the LEGUS team then visually inspects all candidates brighter than $-$6 in the V-band and classifies them into four morphological categories: (1) centrally concentrated clusters with spherically symmetric profiles; (2) clusters with asymmetric radial profiles; (3) multi-peaked clusters with underlying diffuse emission; and (4) non-cluster contaminants, such as background galaxies, stars, bad pixels, or edge artefacts. In this work, we limit our analysis to Class 1 (426), 2 (437), and 3 (413) clusters, for a total sample of 1276; we reduce this to 1275 by removing one cluster that we found to be a duplicate entry in the LEGUS NGC 628c and 628e catalogues, located in the small region where the fields overlap. \autoref{fig:NGC628_cl} shows the mosaic V-band image of NGC~628 overlaid with the cluster positions, coloured by their morphological classification. 

\subsection{Observational Completeness Limits}
\label{ssec:completeness}

Because we intend to forward model the cluster population, we require knowledge of the completeness of the observational catalogue. Specifically, we require knowledge of the function $P_\mathrm{obs}(\mathbf{m})$, which describes the probability that a hypothetical star cluster with a vector of magnitudes $\mathbf{m}$ in the various LEGUS filters would be included in the cluster catalogue. To compute this function, we employ the completeness limits reported by \cite{LEGUS2017}, who carry out artificial cluster tests to derive the completeness of the LEGUS automated catalogue generation procedure in each filter independently. For each filter $F$ and a range of magnitudes $m_F$ in that filter, they determine the probability $P_{\mathrm{obs},F}(m_F)$ that a cluster would be recovered by the automated extraction procedure. In what follows we linearly interpolate these tabulated data to obtain a continuous function $P_{\mathrm{obs},F}(m_F)$ that gives the probability that a cluster of arbitrary magnitude $m_F$ in filter $F$ will be recovered.

To determine the completeness for a cluster with a vector of magnitudes $\mathbf{m}$, we use 10,000 Monte Carlo trials. In each trial, we randomly assign each filter $F$ to be a detection or a non-detection with probability $P_{\mathrm{obs},F}(m_F)$ as determined from the interpolated artificial cluster test results. We then determine from this set of detections and non-detections if the cluster would be catalogued following the same criteria used in construction of the actual LEGUS catalogue, i.e., the cluster is catalogued only if it is (1) detected in V-band and at least three other bands, and (2) has a visual magnitude $M_\mathrm{V} \leq -6$. We then take $P_\mathrm{obs}(\mathbf{m})$ to be equal to the fraction of the Monte Carlo trials in which the cluster is catalogued.

We use this method to calculate the completeness both of our library of synthetic clusters (see \autoref{cha:methods}) and of the actual LEGUS catalogue. For the latter, we find a total of 97 clusters for which we estimate $P_{\mathrm{obs},F}(\mathbf{m}) = 0$; these can be present in the catalogue because a small number of clusters with $m_\mathrm{V} > -6$ were added by hand. To avoid introducing errors in our completeness estimate we remove these from the sample. We remove a total of 97 clusters from the sample of NGC~628 due to their zero completeness values, resulting in a final sample size of \nclus\ clusters.

\section{Methods}
\label{cha:methods}
In this section, we describe the pipeline used to derive the cluster demographics, motivated by the forward Bayesian modeling approach demonstrated in \citet{SLUG2019}. For reasons of brevity we only summarise the method, and refer readers to \citeauthor{SLUG2019} for details.



\subsection{Overview of the method}

Given a set of unresolved photometric measurements of a cluster population, how do we infer their underlying demographics? To answer this question, we first propose a joint distribution of mass $M$, age $T$, and extinction $A_\mathrm{V}$ for the cluster population, which we denote as $f (M, T, A_\mathrm{V}\mid \vtheta)$, where $\vtheta$ represents a vector of parameters describing the joint distribution $f$. For example, if we assume the mass distribution of clusters is described by a Schechter function $f(M, T, A_\mathrm{V}\mid \vtheta) \propto M^{\alpha_M} \exp(-M/M_\mathrm{break})$, then $\vtheta$ contains the slope $\alpha_M$ and break mass $M_\mathrm{break}$ of the Schechter function. We describe the functional forms we consider for $f(M, T, A_\mathrm{V}\mid\vtheta)$ and the parameters they involve in \autoref{ssec:models and priors}.

We seek to derive the posterior distribution of $\vtheta$. We do so in the usual way for a Bayesian method, but writing the posterior as the product of a prior and a likelihood function
\begin{equation}
\label{eq:3.1}
p_\mathrm{post}(\vtheta\mid \{\mathbf{m}\}) \propto \mathcal{L}(\{\mathbf{m}\}\mid\vtheta) p_\mathrm{prior}(\vtheta).
\end{equation}
where $\mathcal{L}(\{\mathbf{m}\} \mid \vtheta)$ is the likelihood for our set of $N_\mathrm{obs}$ photometric measurements. We defer a discussion of priors to \autoref{ssec:models and priors}, and present details of our method of calculating the likelihood function in \autoref{ssec:likelihood}, but to summarise the latter here, we compute the likelihood function using a Gaussian mixture model derived from a large library of synthetic star clusters; as we change $\vtheta$, we adjust the weights applied to this library, which changes the value of $\mathcal{L}(\{\mathbf{m}\}\mid\vtheta)$. Our method can therefore be summarised into the following steps.
\begin{enumerate}
    \item We generate a library composed of synthetic clusters with weights based on a proposed distribution of cluster physical properties and the observational completeness of the survey.
    \item Using the newly created library, we produce a synthetic distribution in photometric space, which will be used to compare with the observations.
    \item We adjust the model parameters $\vtheta$ to maximise the resemblance between the observations and the photometric distribution, as parameterised by the likelihood function. As we do so, we map out the posterior distribution.
\end{enumerate}

We carry out the final step of this procedure using a Markov Chain Monte Carlo (MCMC) method as implemented in the software package \textsc{emcee} \citep{2013emcee}. For all the calculations presented in this paper we use 4000 iterations of 100 walkers, discarding the first 300 iterations for burn-in; both visual inspection of the posteriors and quantitative evaluation of the auto-correlation time indicates that the chains are well-converged. We describe the first two steps of the method -- generating and re-weighting the library, in \autoref{ssec:likelihood}.

\subsection{Calculation of the likelihood function}
\label{ssec:likelihood}

As discussed above, we compute the likelihood function using the Gaussian mixture model described in \citet{SLUG2019}, which operates on a library of $N_\mathrm{lib}$ synthetic clusters, each of which is characterised by a mass $M$, age $T$, extinction $A_\mathrm{V}$, and a vector of photometric magnitudes $\mathbf{m}$ in each of the filters used in the observations. The full library is further described by a vector of photometric bandwidths $\mathbf{h}$, which we set to 0.1 mag in all filters; see \citet{2015Powerlaw} and \citet{SLUG2019} for detailed discussion of the meaning of the bandwidth and the motivation for choosing this value. The synthetic clusters in the library are generated using the Padova tracks including models for asymptotic giant branch (AGB) stars \citep{2005PadovaAGB}, and are integrated with the `starburst99' treatment of stellar atmospheres \citep{Leitherer99a}. Details of the library construction are provided in \aref{app:library}. We note that the choice of stellar tracks does affect some of our results for cluster demographics. We consider both Padova-AGB and MIST \citep{2016MIST} models, and choose to use the former because they yield model luminosity functions that more closely match the observations. See \aref{app:stellar_tracks_changes} for details. To model extinctions, we adopt a Milky Way extinction curve in which the optical and UV (UV, U, B and V bands) extinctions are taken from \citet{1999Fitz}, and the IR (I band) extinctions are obtained from \citet{1984Landini}. 
Given the library, \citet{SLUG2019} show that the likelihood function for the parameters $\vtheta$ can be written as (their equation 12)
\begin{equation}
\label{eq:likelihood}
\mathcal{L}(\{\mathbf{m}\}\mid\vtheta) \propto \prod_{i=1}^{N_\mathrm{obs}} \left[\mathcal{A}(\vtheta)\sum_{j=1}^{N_\mathrm{lib}} w_j(\vtheta) \mathcal{N}(\mathbf{m}_i \mid \mathbf{m}_j, \mathbf{h}'_i)\right],
\end{equation}
where $w_j(\vtheta)$ is the statistical weight of the $j$th library cluster (which depends on the parameters $\vtheta$ as described below), $\mathbf{m}_i$ and $\mathbf{m}_j$ are the vectors of magnitudes for the $i$th observed and $j$th library clusters, respectively, $\mathcal{A}(\vtheta) = [\sum_{j=1}^{N_\mathrm{lib}} w_j(\vtheta)]^{-1}$ is a normalisation factor, and $\mathcal{N}(\mathbf{x}\mid\mathbf{x}_0, \boldsymbol{\sigma})$ is the standard multidimensional Gaussian distribution centred at $\mathbf{x}_0$ and with width $\boldsymbol{\sigma}$, evaluated at position $\mathbf{x}$. The quantity $\mathbf{h}_i'$ is given by $\mathbf{h}'_i = (\mathbf{h}^2 + \boldsymbol{\sigma}_i^2)^{1/2}$, where $\boldsymbol{\sigma}_i$ the observational error on the magnitude of cluster $i$; the quantities $\mathbf{h}'_i$, $\mathbf{h}$, and $\boldsymbol{\sigma}_i$ are vectors with one element per filter, and the expression for $\mathbf{h}'_i$ should be understood as applying separately to each filter. If the magnitude of a specific filter of the cluster is undetected, the product of the observed and library cluster magnitudes becomes zero, thus not contributing to the likelihood function. In practice we evaluate \autoref{eq:likelihood} numerically using the \textsc{cluster\_slug} module of the \textsc{slug} software suite \citep{2015SLUG}, which implements a tree-based order $N_\mathrm{obs}\ln N_\mathrm{lib}$ algorithm for performing the calculation that is much faster than a naive brute force evaluation, which would have a computational cost of order $N_\mathrm{obs} N_\mathrm{lib}$. The \textsc{cluster\_slug} module and the entire \textsc{slug} \citep{2012SLUG, 2014slug, 2015SLUG} software suite are freely available from \href{http://www.slugsps.com}{http://www.slugsps.com}.

The quantity in square brackets in \autoref{eq:likelihood} is the distribution of photometric magnitudes for the cluster library evaluated at the vector of magnitudes $\mathbf{m}_i$ for the $i$th observed cluster. This distribution, and thus the likelihood function as whole, depends on the parameters describing the cluster population $\vtheta$ only through the weight functions $w_j(\vtheta)$ that describe the statistical weight of each library cluster. The relationship between weights and $\vtheta$ is given by (equation 9 of \citealt{SLUG2019})
\begin{equation}
w_j(\vtheta) = P_\mathrm{obs}(\mathbf{m}_j) \frac{f(M_j, T_j, A_{\mathrm{V},j} \mid \vtheta)}{p_\mathrm{lib}(M_j, T_j, A_{\mathrm{V},j})}.
\label{eq:weights}
\end{equation}
The denominator $p_{\mathrm{lib}(M_j, T_j, A_{\mathrm{V}, j})}$ is the distribution function describing the sampling density of the library (see \hyperref[app:library]{Appendix~\ref*{app:library}}), while the numerator $f(M_j, T_j, A_{\mathrm{V}, j} \mid \vtheta)$ is distribution of the physical properties given $\vtheta$, and the pre-factor $P_\mathrm{obs}(\mathbf{m}_j)$ is the probability that a cluster with vector of magnitude $\mathrm{m}_j$ would be included in the LEGUS catalogue. The terms in \autoref{eq:weights} can be intuitively understood as follows. The factor $P_{\mathrm{obs}}(\mathbf{m}_j)$ accounts for the fact that only a fraction of clusters with magnitudes $\mathbf{m}_j$ will be observed due to the completeness limits; we compute this probability as described in \autoref{ssec:completeness}. The denominator is the probability density for drawing clusters with a particular combination of physical parameters $(M_j, T_j, A_{\mathrm{V}, j})$ while constructing the library. Finally, the numerator represents the true probability density for a cluster population described by the parameter set $\vtheta$, so that the ratio $f(M_j, T_j, A_{\mathrm{V},j} \mid \vtheta) / p_{\mathrm{lib}}(M_j, T_j, A_{\mathrm{V},j})$ represents the factor by which we must up- or down-weight the library so that clusters in the library have the same mass, age, and extinction distribution as clusters in reality; if this weight factor is unity, then our library is sampled from the same distribution of cluster properties as the real population. 

Intuitively, then, our method consists of iteratively adjusting the parameters $\vtheta$ and thus the weights $w_j(\vtheta)$ to bring the predicted photometric distribution into as close agreement as possible with the observed one. This will in turn adjust the mass and age distributions, since these are determined by the same vector of parameters $\vtheta$ as the luminosity distribution. Again, we remind readers that this is just an intuitive description of the underlying process; formal proof that \autoref{eq:likelihood} is the correct likelihood function to accomplish this adjustment, along with some details of how we handle complications like clusters where some filters are missing due to the fields of view in the different filters not being perfectly overlapping, is provided in \citet{SLUG2019}.

\subsection{Demographic models and priors}
\label{ssec:models and priors}

Cluster demographics depend on cluster formation and destruction mechanisms, and thus the set of candidate parametric model distributions $f(M,T,A_\mathrm{V}\mid\vtheta)$ we consider is necessarily informed by theoretical expectation. In this work, we consider the two most prominent models: mass-independent destruction (MID; \autoref{sssec:mid}; \citealt{2005MID,2009MID, 2015MID}) and mass-dependent destruction/disruption (MDD; \autoref{sssec:mdd}; \citealt{2005MDD, 2007MDD, 2021MDD}). We couple both to a parametric model for the distribution of extinctions (\autoref{sssec:av_dist}).

\subsubsection{The mass-independent disruption (MID) model}
\label{sssec:mid}

In the MID model, the rate at which star clusters are destroyed is independent of cluster mass. The cluster mass function therefore has the same shape at all ages, and the distribution function $f(M,T,A_\mathrm{V}\mid\vtheta)$ can be separated into two distinct functions, one describing the mass distribution $p_M(M)$ and one the age distribution $p_T(T)$. Given the observational evidence that the mass function is a (possibly) truncated powerlaw, we will adopt a functional form for $p_M(M)$ given by a Schechter function,
\begin{equation}
p_M\left(M\right) \propto M^{\alpha_{M}} \exp \left(-\frac{M}{M_{\mathrm{break}}}\right).
\label{eq:schechter}
\end{equation}

For the age distribution, photometry cannot differentiate between a bound an unbound cluster, so the age distribution is required to be flat for times that are shorter than the physical time required for the stars in a cluster to disperse; of course the distribution can also be flat out to older ages if the mechanisms responsible for disruption do not begin until some time after a cluster forms. Regardless of its physical, origin, we call the time at which cluster disruption begins $T_{\mathrm{MID}}$. After this time, clusters will disrupt, and we approximate the age distribution as a powerlaw. We therefore have 
\begin{equation}
p_T(T) \propto \begin{cases}1, & \text{ if }T<T_{\text {MID }} \\ \left(\frac{T}{T_{\text {MID }}}\right)^{\alpha_{T}}, &\text{ if } T>T_{\text {MID }}\end{cases}.
\end{equation}
Thus the joint mass-age distribution in the MID model is
\begin{equation}
\frac{\mathrm{d}^{2} N}{\mathrm{~d} M \mathrm{~d} T} \propto M^{\alpha_{M}} \exp \left(-\frac{M}{M_{\text {break }}}\right) \max \left(T, T_{\text {MID }}\right)^{\alpha_{T}}.
\label{eq:mass_age_mid}
\end{equation}

The MID model therefore has 4 free physical parameters that we place in our vector $\vtheta$: $\alpha_{M}$, $\log M_{\mathrm{break}}$, $\alpha_{T}$, and $\log T_{\mathrm{MID}}$. We adopt flat priors on $\alpha_M$ from $-4$ to 0 and $\alpha_T$ from $-3$ to 0, reflecting a broad range around previous literature values; we will see that these choices have little effect, as our MCMC never approaches these boundaries. The priors on $\log M_\mathrm{break}$ and $\log T_\mathrm{MID}$ require somewhat more thought. As for $\log M_{\mathrm{break}}$, we impose a flat prior from 2 to 7 because with we barely see clusters more massive than $10^{6.5} M_{\odot}$. The lower mass limit of 2 in log scale is to ensure a reasonable MCMC walker range.      For $\log (T_\mathrm{MID}/\mathrm{yr})$ we impose flat priors from 5 to 10 based on physical plausibility. At the lower end, clusters cannot disperse on less than a crossing timescale, and even the densest clusters detected in LEGUS have crossing timescales well above $10^5$ yr. The upper limit is roughly the age of the Universe.

\subsubsection{The mass-dependent disruption (MDD) model}
\label{sssec:mdd}

For the MDD model, clusters lose mass at a rate that varies as a pure powerlaw function of their current mass \citep{2007MDD,2021MDD}, $\mathrm{~d}M/\mathrm{~d}T \propto -M^{\gamma_{\mathrm{MDD}}}$. For such a mass loss rate, the present-day mass of a cluster born with initial mass $M_i$ at age $T$ is
\begin{equation}
    \label{MDD eq}
    M=M_{\mathrm{i}}\left[1-\gamma_{\mathrm{MDD}}\left(\frac{M_0}{M_{\mathrm{i}}}\right)^{\gamma_{\text {MDD }}} \frac{T}{T_{\mathrm{MDD}, 0}}\right]^{1 / \gamma_{\text {MDD }}},
\end{equation}
where $T_{\mathrm{MDD}, 0}$ represents the time required for a cluster with mass $M_0$ to have fully disrupted, defined as having reached a present-day mass $M=0$. The joint mass-age distribution therefore obeys
\begin{equation}
    \frac{\mathrm{d}^{2} N}{\mathrm{~d} M \mathrm{~d} T} \propto \frac{\mathrm{d}^{2} N}{\mathrm{~d} M_{\mathrm{i}} \mathrm{~d} T} \frac{\mathrm{d} M_{\mathrm{i}}}{\mathrm{~d} M}.
\end{equation}
If the distribution of initial masses $M_i$ follows the Schechter function form given by \autoref{eq:schechter}, then we have
\begin{equation}
    \frac{\mathrm{d}^{2} N}{\mathrm{~d} M \mathrm{~d} T} \propto M^{\alpha_{M}} \eta^{\alpha_{M}+1-\gamma_{\text {MDD }}} \exp \left(-\eta \frac{M}{M_{\text {break }}}\right),
    \label{eq:mass_age_mdd}
\end{equation}
where
\begin{equation}
    \eta(M, T) = \left[1+\gamma_{\mathrm{MDD}}\left(\frac{M_0}{M}\right)^{\gamma_{\text {MDD }}} \frac{T}{T_{\mathrm{MDD}, 0}}\right]^{1 / \gamma_{\text {MDD }}}
\end{equation}
is the ratio of the initial and present-day cluster masses for a cluster of present-day mass $M$ and age $T$.

The MDD model therefore has four free parameters: $\alpha_M$, $\log M_\mathrm{break}$, $\gamma_\mathrm{mdd}$, and $T_{\mathrm{MDD,0}}$; note that $M_0$ is not a separate parameter, because the mass-age distribution depends only on the combination of parameters $M_0^{\gamma_\mathrm{MDD}}/T_\mathrm{MDD,0}$. We therefore without loss of generality choose $M_0 = 100$ M$_\odot$ in what follows. However, for reader convenience we will also report the commonly-used $t_4$ parameter, which is simply the disruption time for a cluster of mass $10^4$ M$_\odot$; this is given by $t_4 = 10^{2\gamma_\mathrm{MDD}} T_\mathrm{MDD,0}$. We adopt the same priors on $\alpha_M$ and $\log M_\mathrm{break}$ as in the MID model (\autoref{sssec:mid}). For $\gamma_\mathrm{MDD}$, previous observational estimates and $N$-body simulations give values in the range $0.6-0.7$ \citep{2010Lamers}, and we adopt broad priors that include this range: we take $\gamma_\mathrm{MDD}$ to be flat from 0 to 1. These limits stem from physical considerations: if $\gamma_\mathrm{MDD} < 0$ then low-mass clusters lose mass more slowly than massive ones, contrary to the physical expectations of the model, while if $\gamma_\mathrm{MDD} \geq 1$ then no clusters ever disrupt because there is no $T$ for which $M = 0$. Finally, we adopt flat priors on $\log(T_\mathrm{MDD,0}/\mathrm{yr})$ from 4 to 10; these limits are broad enough not to matter, because none of our walkers ever approach them.

\subsubsection{Dust extinction}
\label{sssec:av_dist}

The exact functional shape of the distribution of dust extinctions $p_{A_\mathrm{V}}(A_\mathrm{V})$ is unknown, so we model it as non-parametrically as possible. Following \citet{SLUG2019}, we adopt a simple piece-wise linear form over the range $A_\mathrm{V}= 0-3$ mag characterised by six free parameters ($p_{A_{\mathrm{V}, i}}, 0\leq i \leq 6$) to be fit, representing the value of the PDF at $A_{\mathrm{V}}$ = $i\,\Delta A_\mathrm{V}$ mag with $\Delta A_\mathrm{V} = 0.5$ mag. Thus the functional form we adopt for the extinction distribution is
\begin{eqnarray}
    \lefteqn{
    p_{A_\mathrm{V}}(A_\mathrm{V}) \propto
    }
    \\
    & &
    \begin{cases}
    p_{A_\mathrm{V},i} + \left(p_{A_\mathrm{V},i+1} - p_{A_\mathrm{V},i}\right)\left(\frac{A_\mathrm{V}}{\Delta A_\mathrm{V}}-i\right), &
    i < \frac{A_\mathrm{V}}{\Delta A_\mathrm{V}} \leq i+1 \\
    0, & A_\mathrm{V} > 3\,\mathrm{mag}
    \end{cases}
    \nonumber
\end{eqnarray}
We treat $p_{A_{\mathrm{V}, i}}$ for $i = 0 - 5$ as parameters of our model to be fit, with $p_{A_\mathrm{V},6}$ fixed by the requirement that the total area under the PDF be unity. We set the priors on $p_{A_{\mathrm{V}, i}}$ to be flat for all values $> 0$, subject to the requirement that $p_{A_{\mathrm{V}, i}} > 0$ remain positive for all $A_\mathrm{V}$. 

Combining this with the mass and age distributions, our final functional form to be fit is
\begin{equation}
    f(M,T,A_\mathrm{V}\mid\vtheta) \propto \frac{\mathrm{d}^{2} N}{\mathrm{~d} M \mathrm{~d} T} \, p_{A_\mathrm{V}}(A_\mathrm{V}),
\end{equation}
with $\mathrm{d}^2N/\mathrm{d}M\mathrm{~d}T$ given by \autoref{eq:mass_age_mid} or \autoref{eq:mass_age_mdd} for the MID or MDD models, respectively.

\subsection{Model selection using Akaike weights}

The method described thus far allows us to compute the posterior PDFs of the model parameters $\vtheta$ for both the MID and MDD models. To determine whether the MID or MDD model provides a better fit and more accurate description of the data, we use the Akaike information criterion (AIC) to assess the fit quality provided by non-nested models. For models tested using AIC, we first identify the walker with the highest likelihood, then compute,
\begin{equation}
    \mathrm{AIC}_{(\text {MID,MDD})}=2 k-2 \ln \hat{\mathcal{L}}_{\text {(MID,MDD)}},
    \label{eq:AIC}
\end{equation}
where $k$ represents the number of free parameters in the models, and $\hat{\mathcal{L}}$ is the maximum of the likelihood function. In our case, $k$ is 11 for both the MID and MDD models, with four parameters describing the joint mass-age distribution ($\alpha_M$, $M_\mathrm{break}$, and either $\alpha_T$ and $T_\mathrm{MID}$ or $\gamma_\mathrm{MDD}$ and $T_\mathrm{MDD,0}$, six parameters representing the dust extinction shape, and one extra parameter to describe the total number of clusters present in the galaxy. 

For the model comparison, we calculate the Akaike weights for either the MID or MDD model as
\begin{equation}
    w_{\mathrm{MID/MDD}} = \frac{\frac{e^{-\Delta_{\mathrm{MID/MDD}}}}{2}}{\frac{e^{-\Delta_{\mathrm{MID}}}}{2} + \frac{e^{-\Delta_{\mathrm{MDD}}}}{2}}.
    \label{eq:Akaike_w}
\end{equation}
The AIC measures the amount of information in the data preserved by a given model., with the relative Akaike weight $w$ of one model indicating the confidence level at which we can claim that it preserves more information than the other models considered. The model with the highest Akaike weight is our best fit and thus the preferred model. 
\label{ssec:model_sel}

\section{Results}
\label{cha:results}
In this section, we report our fit parameters and model comparison results (\autoref{ssec:fullcatalogue}), along with comparisons between models and observed photometry to verify that our best-fitting models do a reasonable job at reproducing the observations (\autoref{ssec:photocomparison}). To search for variations in cluster population demographics with galactocentric radius, we also separately analyse clusters in the inner and outer galaxy (\autoref{ssec:radialsplit}).

\renewcommand{\arraystretch}{1.25}
\setlength{\tabcolsep}{2.pt}
\begin{table*}
    \begin{tabular}{ccccccccccc}
            \hline 
            Catalogue &Model& $w$ & $\alpha_{M}$& $\log \left(M_{\mathrm{break}}/M_{\odot}\right) $ & $\alpha_{M_4}$ & $\alpha_{T}$&$\log \left(T_{\mathrm{MID}}/\mathrm{yr}\right)$ & $\log (T_{\mathrm{MDD,0}}/\mathrm{yr})$ &$\gamma_{\mathrm{MDD}}$ & $t_4(\mathrm{Myr})$ \\ \hline
           \multicolumn{11}{l}{\textit{Full Sample}} \\ \hline
           \multirow{2}{*}{All} &MID & 1.00 & $-2.16^{+0.32}_{-0.04}$ & $5.61^{+0.49}_{-1.24}$ & $-2.85^{+0.10}_{-0.05}$ & $-1.41^{+1.19}_{-0.40}$ & $8.22^{+0.07}_{-1.86}$ & $-$ &$-$ &$-$  \\ 
           &MDD & <$10^{11}$ & $-1.40^{+0.07}_{-0.08}$ &$4.00^{+0.05}_{-0.05}$ & $-$ &$-$ &$-$ &$6.65^{+0.10}_{-0.07}$ & $0.98^{+0.02}_{-0.04}$& $268^{+37}_{-29}$  \\ \hline 
           \multicolumn{11}{l}{\textit{Radial Division}} \\ \hline
           \multirow{2}{*}{$\{ r \leq R_{50}\}$} &MID & 0.6027& $-1.97^{+0.51}_{-0.08}$ & $5.07^{+0.34}_{-0.79}$ & $-2.74^{+0.34}_{-0.07}$ & $-1.84^{+1.47}_{-0.75}$ & $8.28^{+0.07}_{-1.90}$ & $-$ &$-$ &$-$ \\ 
           &MDD & 0.3973 & $-2.24^{+0.04}_{-0.04}$ & $7.24^{+0.51}_{-0.57}$ &  $-$ &$-$ &$-$ & $7.61^{+0.19}_{-0.22}$ & $0.40^{+0.10}_{-0.07}$ & $218^{+47}_{-38}$  \\ \hline
           \multirow{2}{*}{$\{ r > R_{50}\}$ } &MID & 0.2390 & $-1.97^{+0.51}_{-0.08}$ & $5.07^{+0.34}_{-0.79}$ & $-2.74^{+0.34}_{-0.07}$ & $-1.84^{+1.47}_{-0.75}$ & $8.28^{+0.07}_{-1.90}$ & $-$&$-$ &$-$   \\ 
           &MDD & 0.7610  & $-1.79^{+0.10}_{-0.10}$ & $4.47^{+0.25}_{-0.12}$ &  $-$ &$-$ &$-$ & $6.83^{+0.44}_{-0.23}$ & $0.85^{+0.11}_{-0.23}$ & $238^{+62}_{-47}$  \\ \hline

           \multirow{2}{*}{$\{ r\leq R_c \}$ } &MID &0.9175& $-2.18^{+0.04}_{-0.03}$ & $7.20^{+0.56}_{-0.71}$ & $-2.73^{+0.04}_{-0.05}$ & $-1.13^{+0.20}_{-0.22}$ & $8.07^{+0.10}_{-0.16}$ & $-$ & $-$ &$-$  \\ 
           &MDD & 0.0825 & $-2.17^{+0.03}_{-0.03}$ & $7.40^{+0.41}_{-0.54}$ &  $-$ &$-$ &$-$ & $8.06^{+0.13}_{-0.16}$ & $0.22^{+0.07}_{-0.05}$ & $288^{+50}_{-40}$  \\ \hline
           \multirow{2}{*}{$\{ r >     R_c \}$ } &MID &0.8131& $-2.04^{+0.63}_{-0.38}$ & $5.00^{+2.28}_{-0.81}$ & $-2.78^{+0.53}_{-0.35}$ & $-0.75^{+0.47}_{-1.47}$ & $7.74^{+1.66}_{-1.68}$ & $-$ & $-$ &$-$   \\ 
           &MDD &0.1869 &$-2.23^{+0.38}_{-0.27}$ & $5.47^{+1.68}_{-1.08}$ &  $-$ &$-$ &$-$ & $7.96^{+1.47}_{-1.19}$ & $0.61^{+0.29}_{-0.42}$ & $685^{+15845}_{-582}$  \\ \hline

        \multicolumn{11}{l}{\textit{Subpopulations}} \\ \hline
        \multirow{2}{*}{$\{ p_{\mathrm{obs}} \geq 90 \%  \} $} &MID & 1.00 & $-2.16^{+0.17}_{-0.04}$ & $5.32^{+0.35}_{-0.57}$ & $-2.90^{+0.06}_{-0.04}$ & $-1.34^{+1.17}_{-0.65}$ & $8.22^{+0.08}_{-1.85}$ & $-$ & $-$  &$-$  \\ 
           &MDD & < $10^{-4}$ & $-2.14^{+0.08}_{-0.05}$ & $4.91^{+0.15}_{-0.21}$ &  $-$ &$-$ &$-$ & $7.11^{+0.56}_{-0.20}$ & $0.85^{+0.11}_{-0.29}$ & $450^{+141}_{-96}$ \\ \hline
        \multirow{2}{*}{Class 3 Excluded} &MID &  1.00 & $-1.99^{+0.16}_{-0.08}$ & $5.16^{+0.34}_{-0.76}$ & $-2.76^{+0.08}_{-0.06}$ & $-1.05^{+0.98}_{-0.48}$ & $8.21^{+0.11}_{-2.37}$ & $-$ &$-$ &$-$  \\ 
           &MDD & < $10^{-4}$  & $-1.74^{+0.06}_{-0.06}$ & $4.90^{+0.08}_{-0.09}$ &  $-$ &$-$ &$-$ & $6.19^{+0.06}_{-0.05}$ & $0.98^{+0.02}_{-0.03}$ & $91^{+10}_{-9}$ \\ \hline 

        \multirow{2}{*}{OGCs Excluded} &MID & 0.0364 & $-1.50^{+0.07}_{-0.34}$ & $4.26^{+0.19}_{-0.06}$ & $-2.44^{+0.07}_{-0.30}$ & $-0.40^{+0.04}_{-1.01}$ & $6.39^{+1.76}_{-0.05}$ & $-$ &$-$ &$-$  \\ 
           &MDD & 0.9636 & $-1.79^{+0.06}_{-0.05}$ & $4.33^{+0.07}_{-0.06}$ &  $-$ &$-$ &$-$ & $6.64^{+0.19}_{-0.11}$ & $0.93^{+0.05}_{-0.10}$ & $217^{+32}_{-26}$ \\ \hline 
        \multirow{2}{*}{$\{D^5_{\mathrm{norm}} > 2\}$} &MID &1.00 & $-2.17^{+0.33}_{-0.04}$ & $5.58^{+0.39}_{-1.21}$ & $-2.86^{+0.11}_{-0.04}$ & $-1.52^{+1.31}_{-0.70}$ & $8.24^{+0.08}_{-1.91}$ & $-$ &$-$ &$-$  \\ 
           &MDD &< $10^{-11}$ & $-2.16^{+0.05}_{-0.04}$ & $5.42^{+0.39}_{-0.24}$ &  $-$ &$-$ &$-$ & $8.11^{+0.43}_{-0.46}$ & $0.27^{+0.25}_{-0.23}$ & $400^{+78}_{-58} $\\ \hline 


        \end{tabular}
    \caption{Fitting results. The first column indicates the subset of the LEGUS catalogue used in the fit, the second indicates the type of model fit (MID or MDD), the third gives the Akaike weight for that model, and the remaining columns give marginal posterior PDFs derived for each model parameter, reported as ${q_{50}}^{+(q_{84}-q_{50})}_{-(q_{50}-q_{16})}$, where $q_N$ denotes the estimated $N^{\mathrm{th}}$ percentile. Thus our central values are the median of the PDF, and the ranges shown correspond to the 68\% confidence interval. The co-rotation radius is represented by $R_c$, while the median galactocentric radius is denoted as $R_{50}$. In the subpopulation section, we present trimmed catalog fits based on four different trimming criteria. Firstly, we consider the cluster population denoted by ${p_{\mathrm{obs}} \geq 90\%}$, which includes clusters with completeness values exceeding 90\%. Additionally, we provide three additional trimmed catalog fits where we exclude class 3 objects, old globular clusters, and clusters with a normalized distance to the nearest $5^{\mathrm{th}}$ neighbor in the library greater than $2\sigma$.}

        \label{Tab:params}
\end{table*}

\subsection{Full Catalogue Fits}
\label{ssec:fullcatalogue}

We first analyse the full LEGUS NGC 628 catalogue using the method described in \autoref{cha:methods}. We summarise the marginal posteriors we derive on all model parameters in the first block in \autoref{Tab:params}. We also report results for the nuisance parameters describing the dust distribution in \hyperref[app:Av]{Appendix~\ref*{app:Av}}. Our model comparison yields $w(\mathrm{MID}) \approx 1$ and $w(\mathrm{MDD}) < 10^{-11}$, suggesting that the MID model does a substantially better job capturing the variations in the data, and we will therefore focus on this as our preferred model from this point forward.

\begin{figure} 
    \centering
    \includegraphics[width=\columnwidth]{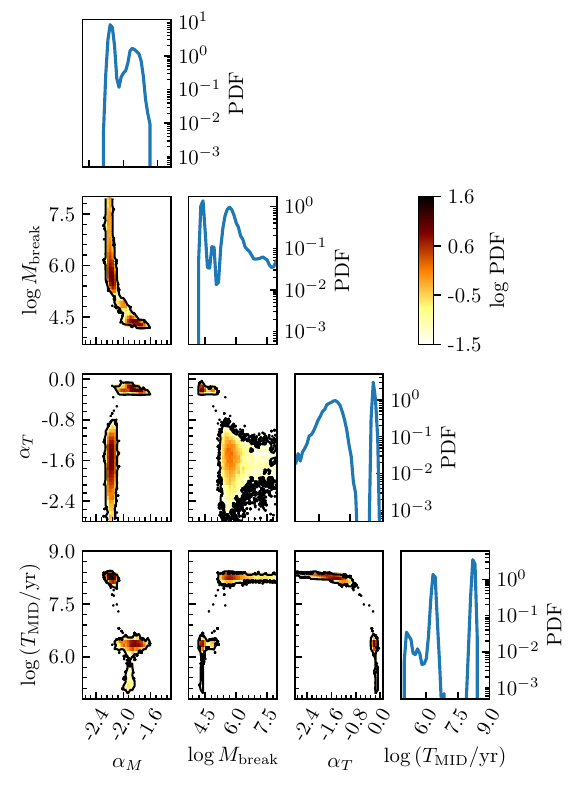}
   \caption{Corner plot showing the one-dimensional and two-dimensional histograms of the posterior PDFs of the parameters $\alpha_M$, $\log {M_{\text{break}}}$, $\alpha_T$, and log $T_{\text{MID}}$ for the full LEGUS NGC 628 catalogue. We omit the nuisance parameters describing the distribution of extinctions $A_V$. The 1D blue histograms show one-dimensional marginal PDFs for each parameter, while the colour maps show log probability densities in various 2D cuts through parameter space. The outermost contour level in the 2D plots is set to enclose 99\% of the samples, and black dots show individual MCMC samples outside this threshold. The extent of each axis is rounded according to the prior range to achieve optimal visualisation. }
    \label{fig:MID_corner}
\end{figure}

\autoref{fig:MID_corner} shows the posterior PDFs of $\alpha_M$, $\log M_{\mathrm{break}}$, $\alpha_T$, and $\log T_{\mathrm{MID}}$ we obtain from our fit. In this plot, the extent of the axes reflects the full range of model parameters allowed by our priors, so we can immediately read off where parameters are well-constrained by the data, versus where they are unconstrained and occupy the full range of values allowed by our priors. These figures have a
few noteworthy features. First, they suggest that cluster disruption in
NGC 628 is weak to non-existent out to ages of $\approx 200$ Myr -- the posterior PDF suggests that either $T_\mathrm{MID} \approx 2\times 10^8$ yr and that disruption becomes fairly rapid only after this point ($\alpha_T \approx -1.4$) or that disruption begins early, $T_\mathrm{MID} \approx 2$ Myr but is very mild ($\alpha_T \approx -0.3$); in either scenario, disruption is limited at the young to moderate ages where our sample has most of its statistical power. Second, the fits 
provide evidence that the mass function is better described by either a truncated Schechter-form CMF with a slope close to the commonly-found $\alpha_M = -2$ but then an exponential cutoff at $M_\mathrm{break} \approx 10^{4.5}$ M$_\odot$ or by a steeper powerlaw with $\alpha_M \approx -2.3$.

\subsection{Photometric Comparisons}
\label{ssec:photocomparison}

Before accepting the results of our fits, we must validate that our models adequately recover the observed luminosity distribution, since matching this distribution is the goal of our forward model.

\subsubsection{Luminosity Functions}
\label{ssec:lum_func}

To visualise the comparisons between the distributions of observations and the library, we first plot the 1D cluster luminosity distributions in the five LEGUS bands, showing both the measured distribution and the luminosity distribution we predict using the model parameters. The model prediction follows immediately from our expression for the likelihood function in terms of our Gaussian mixture model (\autoref{eq:likelihood}), and is simply
\begin{equation}
    p(m) \propto \sum_{j=1}^{N_\mathrm{lib}} w_j(\vtheta) \mathcal{N}(m \mid m_j, h),
    \label{eq:lumdist}
\end{equation}
where $m$ is the magnitude in the filter of interest, $m_j$ is the magnitude of the $j^{\mathrm{th}}$ library cluster in that filter, and $h$ is the library bandwidth. For the observations, we marginalise over the observational scatter using a bootstrap resampling method. Specifically, the LEGUS catalogue reports a central value and an uncertainty (assumed to be Gaussian) on the magnitude of each cluster in each photometric band; to generate the observed distributions, we draw 20,000 samples from this Gaussian for each cluster, and generate an observed luminosity distribution for each realisation. We then plot the median and $5^\mathrm{th}$ to $95^\mathrm{th}$ percentile range of these 20,000 trails.\footnote{We caution that near the completeness limit of the LEGUS sample this resampling procedure is subject to Malmquist bias. The nature of the bias is as follows: for clusters near the $V=-6$ mag lower limit on the LEGUS catalogue, there is a significant chance in any given realisation that they will wind up being assigned a $V$ magnitude fainter than $-6$, leading to them being excluded from the catalogue we produce for that realisation. If the original LEGUS catalogue from which we started included clusters fainter than $V=-6$ mag this would be mostly compensated by clusters just below the magnitude limit of the original catalogue scattering upward in some realisations -- that is, for every cluster with $V = -6.001$ mag in the original LEGUS catalogue that scatters below the magnitude cut in almost half the realisations, there should be another cluster with $V = -5.999$ mag that is excluded from the original catalogue but scatters above the magnitude cut in almost half the realisations. However, because clusters for which the central estimate for $V=-5.999$ mag do not appear in the LEGUS catalogue at all, this compensation does not happen, and as a result we end up under-counting clusters near the magnitude limit by as much as a factor of $\approx 2$. This effect is significant for magnitudes within a few $\sigma$ of the catalogue cutoff, and since the typical photometric error in LEGUS is $\sigma \approx 0.1$ mag, we expect significant bias for $V\gtrsim -6.3$ mag.}

\begin{figure}
    \centering
    \includegraphics[width=0.99\columnwidth]{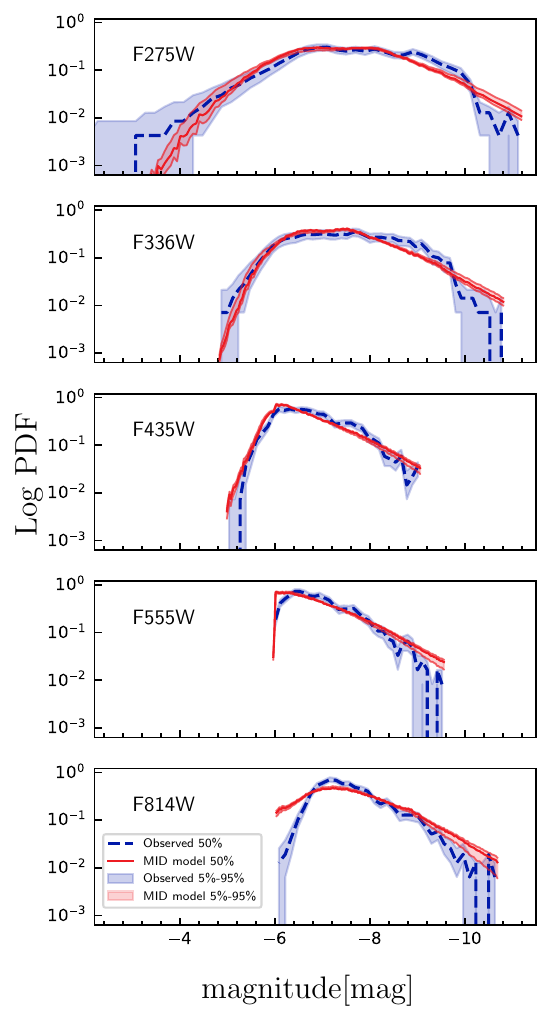}
    \caption{Comparison between observed and model-predicted luminosity functions in UV, U, B, V, I bands. Blue dashed lines show the luminosity distribution of the LEGUS catalogue, divided into uniformly spaced bins; distributions are normalised to have unit integral, and blue shaded regions show the $5^{\mathrm{th}}$ - $95^{\mathrm{th}}$ percentiles in normalised bin counts derived from 20,000 Monte Carlo re-samplings of the observations -- see main text for details of the procedure. The red solid lines illustrate the predicted photometric distributions of the best MID models, computed using the $50^{\mathrm{th}}$ percentile values post burn-in. The shaded bands surrounding these lines indicate the $5^{\mathrm{th}}$ - $95^{\mathrm{th}}$ percentiles for each bin. The blue dashed lines show the $50^{\mathrm{th}}$ percentiles of the photometric distributions in each individual bin, using bootstrapped samples of the observed luminosities. The blue shaded bands surrounding these dashed lines indicate the $5^{\mathrm{th}}$ - $95^{\mathrm{th}}$ percentiles for each bin.}
    \label{fig:1D_MID}
\end{figure}

\begin{figure}
    \centering
    \includegraphics[width=0.99\columnwidth]{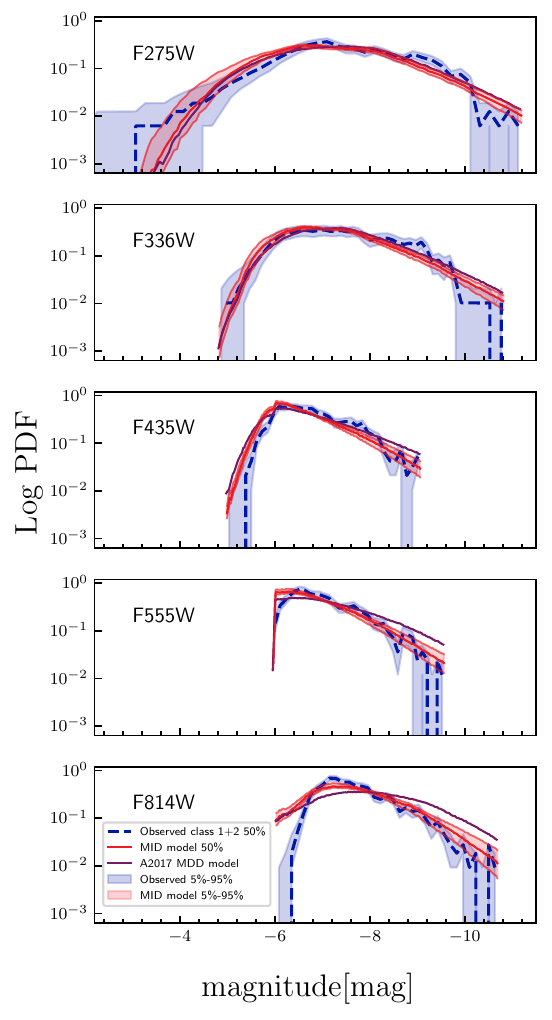}
    \caption{Same as \autoref{fig:1D_MID}, but now showing the luminosity function only for class 1 and 2 clusters, and including a comparison to the luminosity distribution predicted using the best-fitting parameters obtained by \citet{LEGUS2017} (black line).}
    \label{fig:1D_MID_CL12}
\end{figure}

We compare the observed and model-predicted luminosity functions in \autoref{fig:1D_MID}. As the figure shows, our best-fitting model does an excellent job of reproducing the full luminosity function in the bluer bands, and reproduces the shape of the peak of the luminosity function very well in every band. The only place where our model lies significantly outside the observed confidence interval is in the I band, where the observed luminosity function drops off somewhat more steeply than our model fit at both the bright and dim ends. At the dim end this is likely due to our completeness model not being perfectly accurate -- one can see hints of this in the V band as well, where our model cuts off sharply at $-6$ mag, the nominal catalogue limit, but the data fall of slightly more smoothly. The Malmquist bias discussed above may contribute slightly to this discrepancy, but given that the observed luminosity function in $V$ band turns over near $V=-6.5$ mag, roughly $5\sigma$ above the catalogue limit for the typical $\sigma\approx 0.1$ mag uncertainty in LEGUS, it is unlikely to be the dominant effect. At the bright end the discrepancy may be a result of our simple Schechter functional form being too simple to fully capture the full shape fo the mass distribution. A second possibility is that this mismatch is a result of errors in our assumed values of metallicity or nebular covering fraction, a topic we address further momentarily.

It is also interesting to compare the fidelity with which we are able to reproduce the luminosity function with the results of prior studies that derived cluster mass and age distributions from a more traditional backwards-modelling approach rather than our Bayesian forward-modelling method. \citet{LEGUS2017} used the cluster fitting code \textsc{yggdrasil} to assign masses and ages to each cluster in NGC 628, and find that the population is best fit by an MDD model with $t_4 = 190$ Myr, $\gamma_\mathrm{MDD} = 0.65$, $\alpha_M = -2.03$, and $M_\mathrm{break} = 2.03\times 10^5$ M$_\odot$. These results are derived using the same stellar tracks as we use in our library, but come from a fit to the ``exclusive'' catalogue that excludes morphologically-complex class 3 clusters. To compare to these results, we therefore fit the LEGUS catalogue excluding class 3 sources using our method (see \autoref{ssec:robustness} for details), and then generate predicted luminosity functions for both \citet{LEGUS2017}'s best-fit model and our best fit, using the same method as in \autoref{fig:1D_MID}, and compare to the observed LEGUS catalogue excluding class 3 clusters. We show the results in \autoref{fig:1D_MID_CL12}. It is clear that our method performs substantially better at reproducing the observed luminosity function, particularly in the redder bands.

\subsubsection{Colour Distributions}

\begin{figure}
    \centering
    \includegraphics[width=\columnwidth]{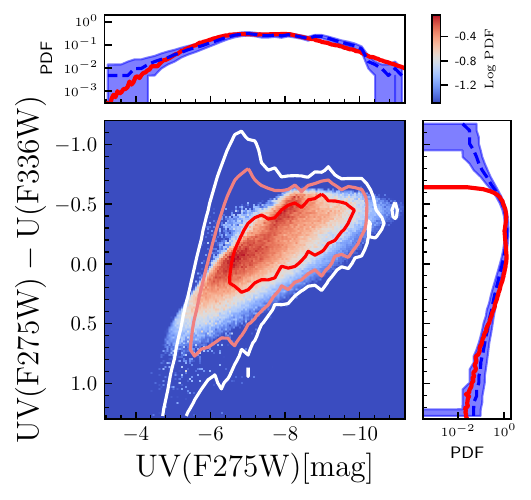}
    \caption{Comparison between observed and model-predicted cluster distributions in UV magnitude versus UV - U colour. The heatmap in the central panel shows the predicted colour-magnitude distribution for our model, and is computed using the $50^{\mathrm{th}}$ percentile values of all parameters. The red, coral, and white contour lines enclose the 50$\%$, 84$\%$, and 95$\%$ of a bootstrap resampling of the observed distribution in the LEGUS catalogue -- see main text for details of the bootstrapping procedure.  The flanking panels compare 1D PDFs of the observed (blue dashed lines) and predicted (red solid lines) colour and magnitude. To visualize the uncertainties around photometric measures, we represent the $5^{\mathrm{th}}$ - $95^{\mathrm{th}}$ percentiles on the observational distribution as blue shaded regions.}
    \label{fig:UV_UVU}
\end{figure}

\begin{figure}
    \centering
    \includegraphics[width=\columnwidth]{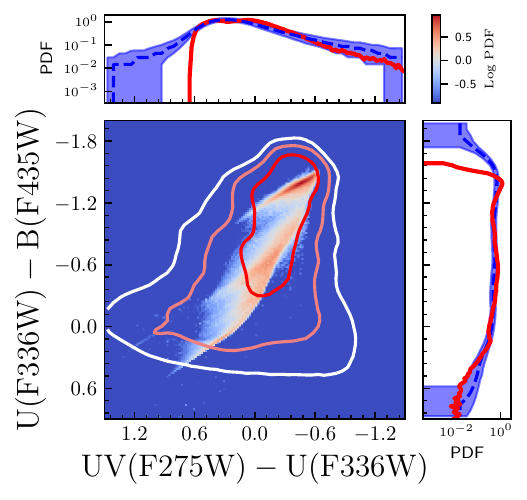}
    \caption{Same as \autoref{fig:UV_UVU}, but showing the U - B versus UV - U colour-colour distributions of the observations and the best-fitting model.}
    \label{fig:UV_U_B}
\end{figure}

 Since our method attempts to fit the full five-dimensional photometric distribution at once, we can also check for agreement between model and observations in multiple dimensions. To this end, we present as examples the UV-U versus U colour-magnitude diagram in \autoref{fig:UV_UVU}, and the UV-U versus U-B colour-colour diagram in \autoref{fig:UV_U_B}; we show the colour-magnitude and colour-colour diagrams for other bands in \hyperref[app:MDDcomp]{Appendix~\ref*{app:MDDcomp}}. As with the luminosity distribution, we compute the colour distribution predicted by our fit directly from the Gaussian mixture model, using an expression analogous to \autoref{eq:lumdist}, and we compute the observed distribution via a Monte Carlo resampling of the LEGUS measurements including their uncertainties. Specifically, to generate the distribution of the observations in colour-magnitude or colour-colour space, we draw 20,000 samples from the Gaussian error distribution for each cluster and band, thereby generating a sample of $\sim 10^7$ points in each possible colour-magnitude and colour-colour diagram. We then plot contour lines that enclose 50\%, 84\%, and 95\% of these samples to indicate the range covered by the observations. These figures also show reasonably good agreement between the observed and model-predicted colour-magnitude and colour-colour distributions, and suggest that our best-fit model provides a reasonable representation of the distribution of observations in 5D photometric space. 

The largest deviations between our model and the observations occur for clusters with the bluest colours. \autoref{fig:UV_UVU} illustrates that the observations include a tail at a UV magnitude of approximately $-6$, featuring the most negative UV-U colour index (bluest colour), which is not well-reproduced by the models. This feature is also present in the U-B colour distributions shown in \autoref{fig:UV_U_B}, where our model exhibits a cutoff in U-B colour space at the bluest end, indicating limitations in reproducing the bluest clusters in the observed samples.

One possible explanation for this discrepancy is that our model library assumes Solar metallicity, while NGC 628 is slightly sub-Solar \citep[e.g.,][]{2010Moustakas}. Consistent with this hypothesis, \citet{2022Deger} compare PHANGS-HST clusters to tracks through colour-space predicted by non-stochastic stellar population synthesis models using a range of metallicity and nebular covering fraction. They find the bluest colours achievable by the models are sensitive to these two parameters, with plausible variations in them having relatively little effect over most of the track, but yielding variations at the extreme blue end by $\sim 0.3$ mag, comparable to the offset we find between our bluest observed clusters and our bluest \textsc{slug} models. However, \citet{2015Powerlaw} show that metallicity variations at this level have relatively little effect when inferring cluster masses or ages. 
In any event, the effects of this discrepancy on population-level statistics should be small, since they involve only a handful of clusters out of our full catalogue of $>1000$. 

While \autoref{fig:1D_MID}, \autoref{fig:UV_UVU}, and \autoref{fig:UV_U_B} allow qualitative comparison between the best-fitting model and the observed photometric distributions, it is also possible to make a quantitative comparison. The conventional approach for assessing whether a model is a good fit typically involves measures such as p-values derived from $\chi^2$ or Kolmogorov-Smirnov \citep[KS;][]{smirnov1939estimation} tests. However, it is important to keep in mind what these tests assess: the p-value indicates the confidence with which we can rule out the null hypothesis that measurement error (for a $\chi^2$ test) or finite sample size (for a KS test) are the \textit{only} sources of disagreement between model and data, i.e., the null hypothesis for a $\chi^2$ or KS test is that the model is perfect and that discrepancies between it and the data arise only because of limitations in the measurement. In our study, we have identified distinct features in the data that are not replicated by any of the models, as demonstrated by the comparisons of colour-colour and colour-magnitude between the models, so we do not expect even our best model to achieve a ``good fit'' in the sense of a high p-value, and thus most of the value in the test is in comparing relative p-values. With this limitation understood, we proceed to carry out KS test comparisons between the observed and model-predicted 1D luminosity distributions shown in \autoref{fig:1D_MID}, and the analogous distributions for the best-fitting MDD model. We present the resulting p-values in \autoref{Tab:ks1}.

As expected, the MID model in the UV band has a p-value high enough to be consistent with the hypothesis that the model is perfect at the few $\sigma$ level. Among the five available bands, the MID model is preferred in all bands except B based on higher p-values, consistent with the finding of \textsc{cluster\_slug} favouring the MID model. However, we caution that one should \textit{not} perform model comparisons just by examining relative p-values, since, among other defects, this approach ignores the higher-dimensional correlations present in the real data, which our comparison based on Akaike weights properly captures. In general we also see that model-observation agreement is better in the bluer bands and worse in the redder ones. Nonetheless, the comparison reinforces the point that, while our models capture the major qualitative features of the observed photometric distributions well, they are clearly not perfect. Given the much larger sample of clusters are we can use compared to previous work, we have the statistical power to detect discrepancies that would have been invisible to earlier analyses on much smaller data sets.


\begin{table}
    \begin{center}
    \begin{tabular}{c c c c c c }
            \hline
            \multicolumn{1}{c}{} & UV & U & B & V & I\\
            \hline
         $\ln p_{\mathrm{MID}}$ & $-1.80$   & $-4.03$ & $-10.04$ & $-3.62$ & $-8.43$ \\
       $\ln p_{\mathrm{MDD}}$ & $-3.09$ & $-4.50$ & $-8.38$ & $-5.13$ & $-11.92$  \\
            \hline
        \end{tabular}
    \caption{Goodness-of-fit log p-values for one-sample KS test comparisons of between the observed and best-fit model luminosity functions (c.f.~\autoref{fig:1D_MID}) in the UV, U, B, V, and I bands, for both the best-fitting MID and MDD models.}
        \label{Tab:ks1}
    \end{center}
\end{table}

\subsection{Inner versus Outer Galaxy Clusters}
\label{ssec:radialsplit}

Having presented the full catalogue fits, we now search for radial variations in cluster demographics. For this purpose we assign every cluster in the LEGUS catalogue a galactocentric radius, taking the centre of NGC 628 to be at $\mathrm{RA}=24.17^\circ$, $\mathrm{DEC}=15.78^\circ$, and adopting an inclination $i=25.2^\circ$ and position angle $\mathrm{PA}=25^\circ$ from North \citep{Grasha2015,Grasha2017}. We use the centre location, inclination, and position angle to deproject the cluster coordinates from RA/DEC to a galactocentric coordinate system. Once clusters have been assigned radii, we divide the catalogue in two ways: first simply by making two sub-catalogues of equal size containing the inner and outer halves of the sample, and second by making two sub-catalogues containing clusters inside and outside the galactic co-rotation radius. The advantage of the former approach is that it ensures that we have equal statistical power in both regions, and thus do not miss radial variations because our inner or outer galaxy sample is too small to see them. The advantage of the latter approach is that co-rotation marks a physically-motivated radius where we might expect to see a change in cluster behaviour, as opposed to the galactocentric radius that marks an equal number division, which is solely a function of the size and location of the LEGUS pointings.

\subsubsection{Equal Number Division}
\label{ssec:eqndiv}
\begin{figure}
    \centering
    \includegraphics[width=\columnwidth]{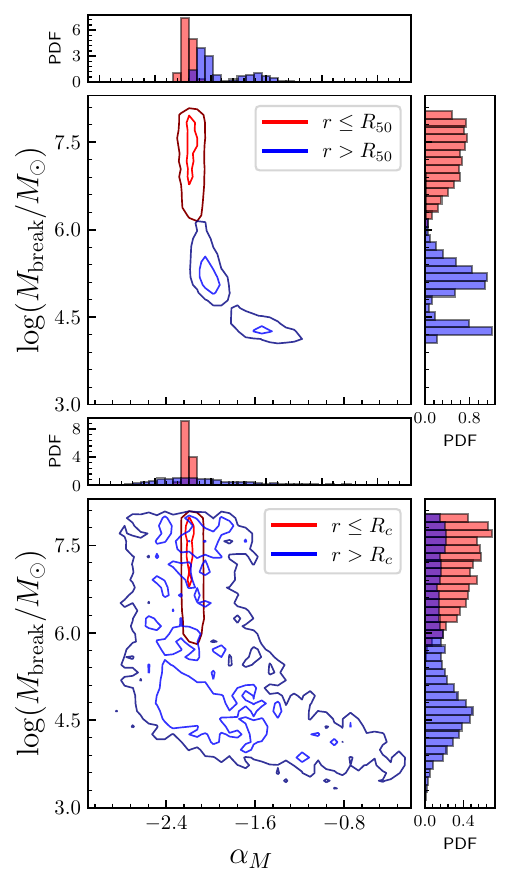}
    \caption{2D marginal posterior joint PDF of the parameters $\alpha_{M}$ and $\log M_{\mathrm{break}}$ that describe the CMF. Upper panel: PDFs for clusters inside (red) and outside (blue) the co-rotation radius $R_c$. Blue and red contour lines in the central panel correspond to loci that enclose 50\% and 95\% of the MCMC samples, and thus correspond approximately to the median and 2$\sigma$ significance confidence level. The histograms flanking the central panel show the corresponding 1D marginal PDFs of $\alpha_{M}$ and $\log M_{\mathrm{break}}$ for the inner (red) and outer (blue) galaxy samples. Lower panel : Same as upper panel, except we show the parameters $\alpha_{M}$ and $\log M_{\mathrm{break}}$ for clusters inside (red) and outside (blue) the median cluster galactocentric radius $R_{50}$}.
    \label{fig:In_out_amMb}
\end{figure}

\begin{figure}
    \centering
    \includegraphics[width=\columnwidth]{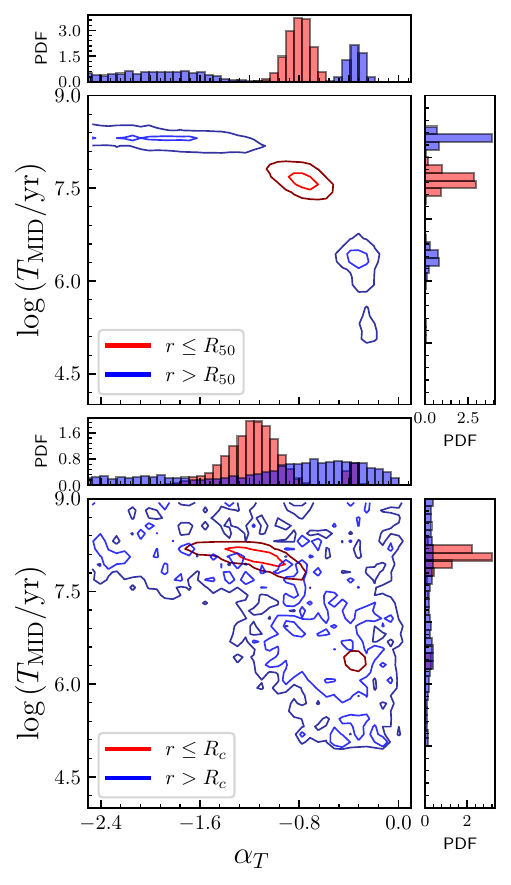}
    \caption{Same as \autoref{fig:In_out_amMb}, expect we show the marginal PDFs of the parameters describing the CAF, $\alpha_{T}$ and $\log T_{\mathrm{MID}}$, rather than the parameters describing the CMF.}
    \label{fig:In_out_aTTmid}
\end{figure}

For our first analysis we divide the LEGUS cluster sample into two radial bins each containing an equal number of clusters. The median radius that makes this even division is $R_{50} = 3.53$ kpc. We fit MID and MDD models to both the subset of clusters with $r \leq R_{50}$ and those with $r > R_{50}$ and report the results the second and third blocks of rows in \autoref{Tab:params}. Our model selection statistics strongly favours the MID model in the inner region, whereas the outer region prefers the MDD model. Comparing these two sub-divided regions, we present the fitted parameters in \autoref{Tab:params} and plotted the posteriors in 2D parameter spaces of the CMF and CAF in \autoref{fig:In_out_amMb} and \autoref{fig:In_out_aTTmid}, respectively. For the outer galaxy age distribution, even though the MDD model is (slightly) preferred, we show results for the slightly less favoured MID model so that we can make a direct comparison.

Examining \autoref{fig:In_out_amMb}, its upper panel shows the marginal posterior PDFs of $\alpha_{M}$ and $M_\mathrm{break}$ of inner and outer regions divided by $R_{50}$.
We notice that the posterior mass distributions of the inner bin (indicated by the red contours) are unimodal, corresponding to a single probability peak in the parameter space centered at $\alpha_{M}\approx -2.2$ and $\log(M_\mathrm{break}/\mathrm{M}_\odot) \gtrsim 6$, while the posterior for the outer region (indicated by blue contours) is bimodally distributed with two distinct probability peaks, one at $\alpha_{M} \approx -2$ and $\log (M_\mathrm{break}/\mathrm{M}_\odot) \approx 5$, the second at $\alpha_{M}\approx -1.5$ and $\log(M_\mathrm{break}/\mathrm{M}_\odot) \sim 4.3$. Despite this difference, however, the $2\sigma$ confidence contours for the two regions slightly overlap, and the 95\% confidence contours clearly lie along a continuous ridge of probability running from low $\alpha_M$ and high $M_\mathrm{break}$ to higher $\alpha_M$ and lower $M_\mathrm{break}$. Thus we formally detect a difference between the inner and outer regions, but at low confidence. To the extent that there is a real difference, it is that there is stronger evidence for a truncation in the mass function in the outer galaxy, an effect consistent with the correlation between truncation mass and star formation rate surface density identified by \citep{2022Wainer}.

The upper panel of \autoref{fig:In_out_aTTmid} provides an analogous plot for the two parameters -- $\alpha_T$ and $T_\mathrm{MID}$ -- that describe the CAF. The qualitative conclusion to be drawn from \autoref{fig:In_out_aTTmid} is similar to that for the mass PDFs as shown in the upper panel of \autoref{fig:In_out_amMb}, i.e., the posterior PDFs of the age parameters for the inner and outer radial bins are mainly separated at $2\sigma$ levels, but are clearly tracing out the same underlying ridge of probability. The contours would overlap if we extended our plots to higher than 95\% confidence; we again therefore consider this a low-confidence detection of a difference. The clusters in the inner region of the galaxy (represented by red contours) begin to disrupt around $T_\mathrm{MID}\approx 30$ Myr, with a relatively rapid disruption rate of $\alpha_T \approx -0.8$ thereafter. Conversely, the PDFs of the age parameters in the outer region (represented by blue contours) appear to be more spread out. The PDF peak for the clusters in the outer region suggests either moderate disruption starting at $1-5$ Myr ($\alpha_T \approx -0.4$), or no disruption until $\sim 200$ Myr. In either case, to the extent that our detection is real, it suggests that disruption always faster in the inner galaxy at ages $\gtrsim 100$ Myr, but that whether disruption in the inner galaxy is faster or slower at younger ages depends on which of the two possibilities identified by the MCMC we favour for the outer galaxy -- for the case of late disruption, $T_\mathrm{mid} \sim 200$ Myr, the inner galaxy has faster disruption at all cluster ages, while for the early but moderate disruption scenario in the outer galaxy, $T_\mathrm{mid} \sim 3$ Myr and $\alpha_T\sim -0.4$, disruption is initially faster in the outer galaxy, but the inner galaxy catches up and overtakes after $\sim 100$ Myr, when $\sim 30-40\%$ of clusters still remain intact.

\subsubsection{Co-rotation Radius Division} 
\label{ssec:cordiv}
An alternative way of separating the cluster population is into those inside and outside the galactic co-rotation radius at $R_c = 6.3$ kpc where the spiral pattern and orbiting clusters move at equal speed \citep{2021CoR}. To the extent that encounters with molecular clouds or other structures associated with the spiral arms influence cluster formation or disruption, we might expect to see changes in cluster population demographics across co-rotation. Making this division, we have 1159 clusters inside $R_c$ and 117 clusters outside $R_c$.

The bottom panels of \autoref{fig:In_out_amMb} and \autoref{fig:In_out_aTTmid} show the inferred posteriors for the clusters located inside and outside the co-rotation radius; we again concentrate on the MID model, because model comparison prefers it to MDD. Here we see differences between the inner and outer galaxy posterior PDFs that are qualitatively consistent with those we observed for a division of the sample into two equal parts. However, the small number of clusters in the outer galaxy sample when we divide at co-rotation ensures that the posterior PDFs are very broad, and, particularly for the age distribution, largely just reflect our priors. Given the very broad outer galaxy posteriors, we cannot rule out the possibility that the cluster demographics inside and outside co-rotation are the same. Doing so would likely require significantly more outer-galaxy clusters than the LEGUS catalogue provides.


\subsection{Individual cluster ages and masses}
\label{ssec:individual}

In addition to the population-level inference of the marginal posteriors, we can also compute individual cluster-by-cluster fits for cluster masses and ages in NGC 628 using the method described in \citet{2015SLUG, 2015Powerlaw}, with the important difference that since we have now fit for the mass and age distribution of NGC 628 we can use this result as an informative prior. We briefly summarise the method here and refer readers to \citet{2015Powerlaw} for a full description. The method is analogous to that described in \autoref{cha:methods}: we consider a single cluster with a vector of magnitudes $\mathbf{m}$ and corresponding uncertainties $\boldsymbol{\sigma}$, and invoke Bayes' theorem to write the posterior probability distribution for the cluster mass $M$, age $T$, and extinction $A_\mathrm{V}$ as
\begin{equation}
    p_\mathrm{post}(\boldsymbol{\psi} \mid \mathbf{m},\boldsymbol{\sigma}) \propto \sum_{j=1}^{N_\mathrm{lib}} w_j(\boldsymbol{\psi}_j) \mathcal{N}(\mathbf{m} \mid \mathbf{m}_j, \mathbf{h}'_j) \mathcal{N}(\boldsymbol{\psi} \mid \boldsymbol{\psi}_j, \mathbf{h}_{\boldsymbol{\psi}}),
    \label{eq:dist_individual}
\end{equation}
where $\boldsymbol{\psi} = (\log M, \log T, A_\mathrm{V})$ is the vector of parameters we wish to infer, and quantities subscripted by $j$ are properties of the $j$th library cluster: $\mathbf{m}_j$ and $\mathbf{h}'_j$ are the vectors of photometric magnitudes and bandwidths (identical to those used in \autoref{eq:likelihood}), $\boldsymbol{\psi}_j$ is the (log mass, log age, extinction) vector, $\mathbf{h}_{\boldsymbol{\psi}}$ is the bandwidth in (log mass, log age, extinction), and $w_j(\boldsymbol{\psi}_j)$ is the weight. The prior enters in the calculation of $w_j$: library clusters whose properties are \textit{a priori} more likely are weighted more highly (e.g., lower mass clusters will be assigned more weight than higher mass ones, since most clusters are low mass; c.f.~Equation 18 of \citealt{2015SLUG}).
The difference between our calculation here and that presented in \citet{2015Powerlaw} is that, whereas \citeauthor{2015Powerlaw} adopted a range of candidate uninformative priors, here we use the best-fitting model from our population-level analysis as our prior.

\begin{figure}
    \centering
    \includegraphics[width=\columnwidth]{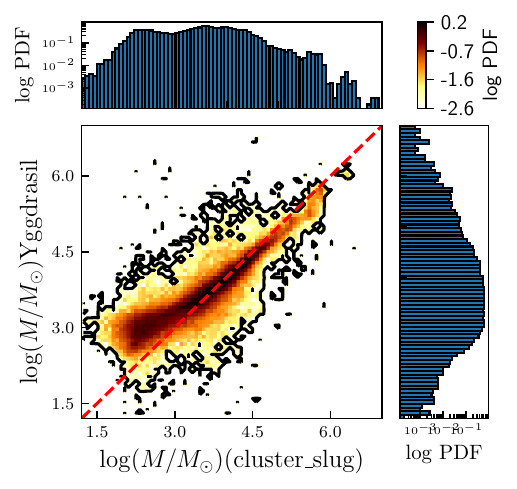}
    \includegraphics[width=\columnwidth]{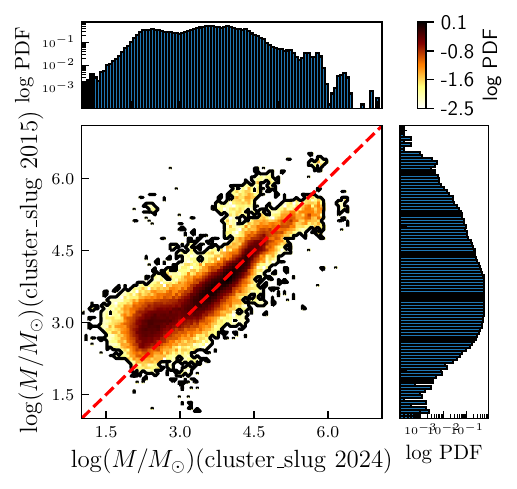}
    \caption{Comparison of NGC 628 cluster mass estimates computed in this paper (\autoref{eq:dist_individual}; horizontal axis) to those computed in alternative ways (vertical axes); the top panel compares to deterministic \textsc{yggdrasil} models (as reported in \citealt{LEGUS2017}), while the bottom panel compares to results derived using the informative priors from \citet{2015Powerlaw}; see main text for details. The 2D density maps display the joint PDF of masses derived by the two methods, estimated using the bootstrap resampling procedure described in the main text (\autoref{ssec:individual}). The top and right flanking panels present the 1D distribution of masses computed by each model.}
    \label{fig:cs_ygg_mass}
\end{figure}

\begin{figure}
    \centering
    \includegraphics[width=\columnwidth]{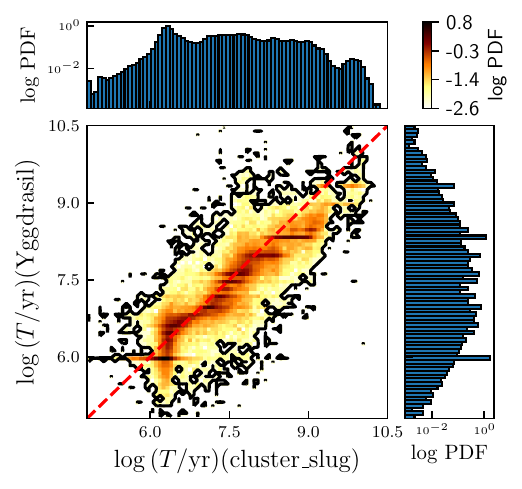}
    \includegraphics[width=\columnwidth]{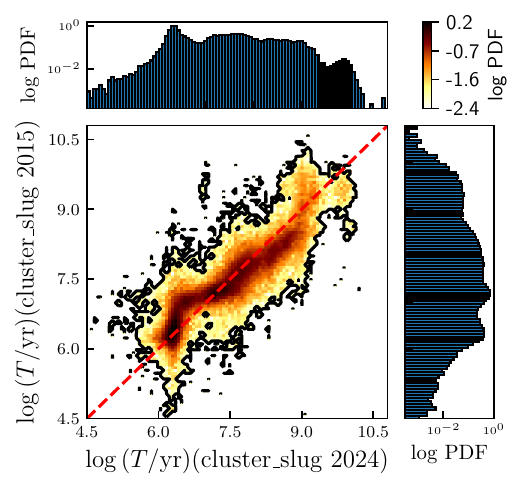}
   \caption{Same as \autoref{fig:cs_ygg_mass}, except we show age estimates instead of mass estimates. Note that the horizontal streaks in the top panel are an artefact of the \textsc{yggdrasil} model, which uses a relatively coarse grid of ages.}
    \label{fig:cs_ygg_age}
\end{figure}

In \autoref{fig:cs_ygg_mass} and \autoref{fig:cs_ygg_age}, we show how the marginal posterior mass and age distributions we derive by this procedure compare to those derived using the deterministic stellar population fitting code \textsc{yggdrasil} \citep{Zackrisson11a} and reported for the clusters in NGC~628 \citep{LEGUS2017} (upper panels), and to those derived using the uninformative priors from \citet{2015Powerlaw} (lower panels); for the latter, we use their $\alpha_M = -1.5$, $\alpha_T = -0.5$ case. We construct these figures using bootstrap resampling: for each cluster in the NGC 628 catalogue, we randomly draw a \textsc{cluster\_slug} mass from the marginal posterior PDF of mass derived from \autoref{eq:dist_individual} using informative priors; to construct the top panel we then draw corresponding masses from the central value and uncertainty (assumed to be Gaussian in $\log M$) for the clusters reported by \citet{LEGUS2017}, while for the bottom panel we also draw the corresponding mass from \autoref{eq:dist_individual}, but with $w(\boldsymbol{\psi}_j)$ evaluated using the uninformative priors. We repeat this procedure until we have a sample of 35,000 pairs for each panel, and we plot the distributions of these points in \autoref{fig:cs_ygg_mass}; the procedure is identical for ages in \autoref{fig:cs_ygg_age}.

Examining these figures, we see that overall \textsc{yggdrasil} tends to produce somewhat higher cluster masses (i.e., the \textsc{cluster\_slug} predicted masses < 1000 $M_{\odot}$ are biased above the one-to-one lines) and younger ages compared to \textsc{cluster\_slug} (i.e., the points are biased slightly below the one-to-one lines), particularly at the lowest masses and youngest ages; agreement is closer at higher mass and age (i.e., the points cluster more closely around the one-to-one lines).\footnote{The horizontal streaks visible in \autoref{fig:cs_ygg_age} are not physically meaningful; they are are artefact of \textsc{yggdrasil} using a discrete, coarsely-sampled grid of ages for its fits.} One explanation for this asymmetry at $M \sim 10^{3}$ M$_\odot$ is, for a given brightness of a cluster, \textsc{yggdrasil} assigns this cluster a unique mass based on a fixed mass-to-light ratio. However, \textsc{cluster\_slug} consider another possibility that it can be a low mass star cluster that happens to contain a massive star that dominates the light profile due to the impartial sampling of the IMF. Because we use an informative prior on mass with a slope $\sim$ -2, which assigns a high probability to less massive clusters, the scenario mentioned above becomes prominent. Consequently, we expect  \textsc{cluster\_slug} to assign systematically lower masses to dim clusters than \textsc{yggdrasil}, as shown in \autoref{fig:cs_ygg_mass}. Similarly, deterministic models interpret a blue colour as requiring a young age, while a stochastic model allows for the possibility that it might also be the result of an older cluster containing a single bright star undergoing a blue loop. Consequently, \textsc{cluster\_slug}'s uncertainty intervals tend to extend to larger masses and ages than \textsc{yggdrasil}'s, and this effect is largest in the least massive and youngest clusters, which are subject to the largest stochastic fluctuations. As noted above, this phenomenon was previously reported by \citet{2015Powerlaw}, so the main new thing to take from our analysis here is that switch from uninformative to informative priors does not appear to alter it significantly.

Consistent with this finding, we see that there are somewhat smaller differences between the \textsc{cluster\_slug} models with the two different priors. Compared to the uninformative prior adopted in \citet{2015Powerlaw}, our population fit has a steeper slope at low mass, since here we have $\alpha_M \approx -2$ as compared to $\alpha_M = -1.5$, and so we again find that the current fit favours somewhat lower masses at the low mass end. Conversely, there are clearly some clusters for which the uninformative prior results in a mass $\gtrsim 10^5$ M$_\odot$, while the informative one reduces the estimated mass to $\approx 10^{4.5}$ M$_\odot$. This is clearly a result of our more sharply truncated mass function.

\begin{figure}
    \centering
    \includegraphics[width=\columnwidth]{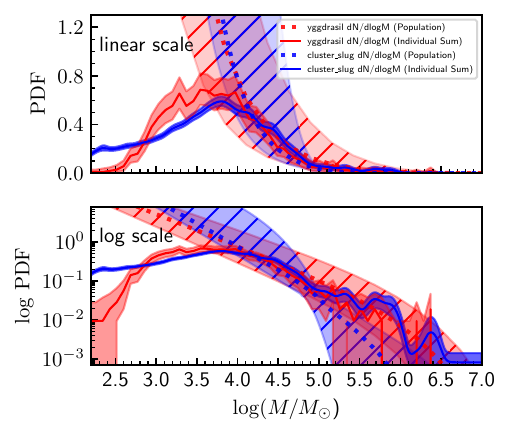}
    \caption{Comparison of the inferred posterior probability density functions (PDFs) in linear (upper panel) and log scales (lower panel) for the mass distribution of the entire population of star clusters using the \textsc{cluster\_slug} and \textsc{yggdrasil} SSP models. The dotted blue line is the mass PDF obtained by summing the marginal mass PDFs derived with \textsc{cluster\_slug} over all clusters; the shaded region around this line is the 5-95\% confidence interval derived from bootstrap resampling via the procedure described in the main text. The red solid line and shaded region show the same quantities derived from \textsc{yggdrasil} deterministic models following \citet{LEGUS2017}. The dashed lines and hatched regions show the corresponding fits to the population mass distribution, computed as described in the main text.}
    \label{fig:mpdfsum_cs_ygg}
\end{figure}

To explore this further, in \autoref{fig:mpdfsum_cs_ygg} we plot the inferred posterior distribution of cluster masses, marginalized over extinction and age, for all the clusters in NGC 628. The blue solid line in this figure is computed by summing the individual cluster posterior PDFs of mass, and the shaded blue band around it is the 5-95\% confidence interval, which we derive from our bootstrap samples; the red solid line and shaded band are the result of applying the same procedure to the \textsc{yggdrasil} results. (We omit the corresponding line for uninformative priors to avoid clutter, but it would show the same qualitative effects.) Comparing these lines, two models agree very well from $10^{4} - 10^{7}$ M$\odot$. At lower masses the  \textsc{cluster\_slug} mass distribution is flatter than the \textsc{yggrdasil} one due to the individual mass PDFs having large dispersions due to \textsc{cluster\_slug}'s inclusion of uncertainties from stochasticity, as described above.

To investigate the consistency between the PDF derived from individual cluster masses and the best-fitting Schechter mass functions we derive, in \autoref{fig:mpdfsum_cs_ygg} we have also over-plotted dashed lines representing those best-fit models; we do so both or the Schechter function derived in this paper, and using the \textsc{yggdrasil}-based provided in \citet{LEGUS2017}; the dotted line is the most likely model, and the hatched region around it is the 5-95\% confidence interval derived from bootstrap resampling of the posterior PDFs. Interestingly, we see that while our best-fitting \textsc{cluster\_slug} model matches the distribution relatively well near the peak of the PDF, it is more sharply truncated at $10^5 M_{\odot}$ than the mass function we get by adding up the individual cluster posterior PDFs. The individual cluster PDFs have a small population of clusters above $10^6$ M$_\odot$, which is absent in our Schechter function fit to the whole population.

This comparison highlights an important feature of our method, which is that it is most sensitive to the peak of the distribution, in contrast to methods based on fitting binned data, which assign equal weight (except for Poisson errors) to nearly unpopulated mass bins as to highly-populated ones. Indeed, there is no \textit{a priori} reason why we should expect to obtain the same result from a method that weights all \textit{clusters} equally versus one that weights all \textit{mass bins} equally, nor why we should expect a forward-modelling method that emphasises reproducing the observed luminosity distribution to match a backward modeling method that has no such constraint. In an ideal world, where our stellar population synthesis models and the functional forms we use to parameterise the cluster population were both accurate, these approaches would agree, but clearly that is not the case given our current models. Nonetheless, this comparison does suggest that in future work, our method might be improved by adopting a more general functional form for the mass distribution that allows independent adjustment of the shape at lower and higher masses, rather than relying on the more prescriptive functional forms (such as the Schechter function) that have traditionally been used. It also means that we should apply the informative priors with caution for very luminous clusters, since our priors may be unfairly discounting the possibility of these being very massive.

\section{Discussion}
\label{cha:discussions}

Here we discuss some implications of our findings. We begin with a series of additional verification checks on our method (\autoref{ssec:robustness}), and then discuss our findings regarding truncation of the cluster mass distribution (\autoref{sec:CMF}) and the cluster age distribution (\autoref{ssec:cluster_disruption}).

\subsection{Method robustness against outliers}
\label{ssec:robustness}

All our results in \autoref{cha:results} are derived from fitting the full LEGUS cluster catalogue. However, there are plausible reasons why one might consider excluding various parts of the catalogue from the analysis, for example because one is worried that particular parts of the catalogue are unreliable, or exercise unreasonable amounts of leverage on the result. For this reason, it is important to assess the robustness of our method  against various potential outlier sub-populations by repeating our analysis on trimmed catalogues where we remove some subset of the clusters. In all these cases, once we remove the target subset of clusters, the remainder of our analysis pipeline is identical to the one used for the full catalogue in \autoref{cha:results}.

\begin{figure}
    \centering
    \includegraphics[width=\columnwidth]{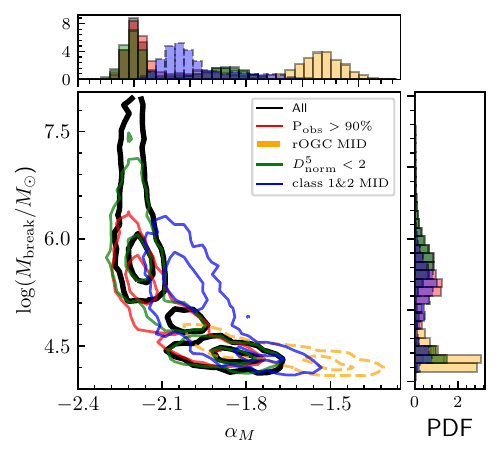}
    \caption{2D marginal posterior joint PDF of the parameters $\alpha_{M}$ and $\log M_{\mathrm{break}}$ that describe the CMF, comparing the results derived from the full LEGUS catalogue (black contours) to those derived for five trimmed catalogues where we have removed clusters with low $P_\mathrm{obs}$ (red, \autoref{sssec:low_pobs}), candidate metal-poor old globular clusters (orange, \autoref{sssec:ogcs}), class 3 clusters (green, \autoref{sssec:class3}), and clusters with $D^5_\mathrm{norm} > 2$ (blue, \autoref{sssec:high_d5}). In all cases the inner contour encloses 50\% of the MCMC samples, and the outer contour encloses and 95\% of the samples. The histograms flanking the central panel show the corresponding 1D marginal PDFs of the full and trimmed catalogues.}
    \label{fig:trimmed_mass}
\end{figure}

\begin{figure}
    \centering
    \includegraphics[width=\columnwidth]{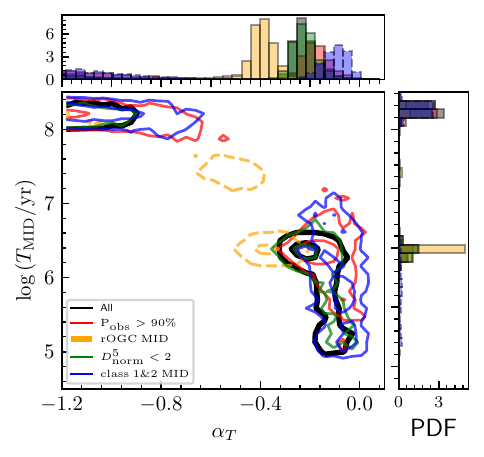}
    \caption{Same as \autoref{fig:trimmed_mass}, except showing the PDF of the parameters $\alpha_{T}$ and $\log(T_{\mathrm{MID}}/\mathrm{yr})$ that describe the CAF for the MID model. For completion, we include the trimmed catalogue comprising only class 1 and 2 clusters as dashed contours. However since this case the catalogue containing only class 1 and 2 clusters that excludes the old globular clusters (rOGC) is better fit by the MDD model, which cannot be directly compared with the MID fits.}
    \label{fig:trimmed_age}
\end{figure}

We compare the mass and age parameters we derive using the reduced catalogues to those we obtain from the full catalogue in \autoref{fig:trimmed_mass} and \autoref{fig:trimmed_age}, respectively, and report the marginal posterior PDFs for all parameters from the reduced catalogues in \autoref{Tab:params}. We also include a table presenting the cluster number statistics of each trimmed catalogue in \autoref{Tab:nc}.

\begin{table}
    \begin{center}
    \begin{tabular}{cccc}
            \hline
            Catalogue& $N_c$ & Best Model& Section\\
            \hline 
            All & 1178 & MID & \autoref{ssec:fullcatalogue}\\
            $\{ p_{\mathrm{obs}} \geq 90 \%  \} $ & 1120 & MID & \autoref{sssec:low_pobs}\\
            OGCs Excluded & 991 & MID &  \autoref{sssec:ogcs}\\
            Class 3 Excluded & 813 & MDD & \autoref{sssec:class3}\\ 
            $\{D^5_{\mathrm{norm}} > 2\}$ & 1160 &MID & \autoref{sssec:high_d5}\\
            \hline
        \end{tabular}
    \caption{Number statistics and best-fitting model for each trimmed catalogue. The last column points to the corresponding section of each trimmed catalogue. $N_c$ represents the number of clusters included in the count that have a non-zero completeness value.}
        \label{Tab:nc}
    \end{center}
\end{table}

\subsubsection{Trimming low-completeness clusters}
\label{sssec:low_pobs}

In \autoref{ssec:completeness} we explain our method to derive the completeness function $P_{\mathrm{obs}}(\mathbf{m})$ for the data and the library. Knowledge of this function, coupled to our forward-modelling techniques, allows us to avoid the traditional method of excluding clusters with masses $\lesssim 5000$ M$_{\odot}$ and ages $\gtrsim 200$ Myr, and in turn allowing us to use essentially all of the LEGUS catalogue, rather than only $\approx 25\%$ of it as in previous work \citep[e.g.,][]{LEGUS2017}. However, our approach is sensitive to our completeness estimates, which therefore represent a potential source of systematic error. To test this possibility, we repeat the numerical experiments with a more conservative approach of only including clusters where we estimate the completeness is $\geq 90\%$. This reduces the sample to 1120 clusters, but ensures that $P_\mathrm{obs}(\mathbf{m})$ never varies by more than 10\%, and thus errors in it have little effect.

Examining \autoref{fig:trimmed_mass} and \autoref{fig:trimmed_age}, we find that the regions of allowed parameter space for both the parameters describing mass ($\alpha_{M}$, $\log M_\mathrm{break}$) and those describing age ($\alpha_T$, $\log T_\mathrm{MID}$) are slightly more concentrated for the high completeness sample, but are almost entirely overlapping with the probability maxima we obtain for the full catalogue; the results are consistent at the $1\sigma$ level. We therefore conclude that possible systematic errors in the completeness function have at most a trivial impact on the outcome.

\subsubsection{Exclusion of candidate metal-poor old globular clusters}
\label{sssec:ogcs}

\citet{2023Whitmore} argues that using a single metallicity for all the clusters in NGC 628 may yield misleading results, because some of the clusters are old globular clusters (OGCs) with substantially lower metallicities than the remainder of the population. Since our method assumes a single (Solar) metallicity, this represents another source of potential bias for us. To assess the robustness of our results against this bias, we re-fit the NGC 628 catalogue after removing candidate OGCs; following \citet{2023Whitmore}, we identify these candidates based on their extremely red colours, $\mathrm{V}-\mathrm{I} > 0.95$ mag.
Examining the comparison between the results derived from this trimmed catalogue and those from the full catalogue in \autoref{fig:trimmed_mass} and \autoref{fig:trimmed_age}, we find that the results derived after removing the OGCs are consistent with those of the full catalogue in the sense that the contours largely overlap, but that exclusion of the OGCs appears to eliminate the possibility of a purely powerlaw but steep mass function that is present in the full catalogue.  Instead, the OGC-cleaned catalogue strongly favours a truncation in the mass function at $M_\mathrm{break} \approx 10^{4.5}$ M$_\odot$. Similarly, of the two possibilities for cluster disruption allowed by the full catalogue -- no disruption to $\approx 200$ Myr and strong disruption thereafter, or weak disruption starting at $\approx 2$ Myr, the OGC-cleaned catalogue favours the latter possibility. We therefore conclude that the possible inclusion of metal-poor clusters in our catalogue, and our approach of fitting the data using a single metallicity, represents a small potential source of systematic error, in that these clusters appear to be responsible for driving our full-catalogue fit toward one of the two possible ``islands'' of probability that it finds. At the same time, however, this effect is not strong enough to skew the results dramatically, in the sense that the posteriors for the OGC-removed catalogue are almost entirely contained within the range of posteriors allowed for the full catalogue.

\subsubsection{Exclusion of class 3 clusters}
\label{sssec:class3}

Our next test is to remove from our catalogue clusters identified as LEGUS class 3. These objects are characterised by irregular, diffuse, or multi-peaked morphologies, likely indicative of a population that is either unbound or is bound but is so young that it has not dynamically relaxed. There is considerable debate in the literature about whether these objects should be categorised as clusters at all (see \citealt{2019ARAA}, and references therein), and previous studies indicate that their mass and age distributions differ from those of the more dynamically-relaxed class 1 and 2 objects \citep{Grasha2015, LEGUS2017}. To investigate the outcome without class 3 objects, we re-fit the NGC 628 catalog after removing them.

Consulting \autoref{fig:trimmed_mass}, the resulting constraints on the mass distribution are only slightly different from those derived from the full catalogue; (\autoref{Tab:params}). The best-fitting break mass is similar to full catalogue fits, and the best-fitting index $\alpha_{M}$ is very slightly shallower, $\approx -2$ rather than $\approx -2.1$. The best-fitting parameters describing cluster ages are nearly identical to those of the full catalogue. Neither of these changes is very significant. Thus we again see that the differences from the full catalogue, while statistically significant, are not particularly physically significant.

\subsubsection{Exclusion of clusters with poor photometric matches}
\label{sssec:high_d5}

In addition to the sub-populations discussed above that have physical meanings, we are also interested in examining potential outliers in photometric space, where our library may not contain any examples capable of reproducing the data. We wish to ensure that these outliers are not biasing our results. To identify photometric outliers, for each catalogue cluster we compute the normalised photometric distance to each library cluster, as defined by \citet{2015SLUG}:
\begin{equation}
    D_{\mathrm{norm},i} \equiv \sqrt{\frac{1}{N_F} \sum_{j=1}^{N_\mathrm{F}} \left( \frac{m_{j, \mathrm{obs}}-m_{j ,i}}{\sigma m_{j,\mathrm{ obs}}}\right)^2},
\end{equation}
where $m_{j, \mathrm{obs}}$ and $\sigma_{j,\mathrm{obs}}$ are the magnitude and uncertainty of the observed cluster in the $j$th of $N_F$ filters, and $m_{j, i}$ is the magnitude of the $i^{\mathrm{th}}$ library cluster in the same filter. We then compute the 5th nearest neighbour distance $D_{\mathrm{norm}}^5$ for each observed cluster, and remove from the catalogue clusters with $D_{\mathrm{norm}}^5 > 2$, i.e., those for which our library includes fewer than 5 cluster that fall within the measured photometric uncertainty interval. This criterion ensures that we exclude clusters that few few or no good matches in our catalogue.

Examining \autoref{fig:trimmed_mass} and \autoref{fig:trimmed_age}, which show the results of fitting the resulting trimmed catalogue, we see that the results align well with the fits from the full sample. The mass parameters are nearly identical, and the age ones differ by less than the uncertainties. This indicates that the method remains robust against outliers in photometric space that exhibit poor correspondence with the library clusters. 

\subsubsection{Summary of robustness tests}

It is worth pausing at this point to summarise the results of our robustness tests. Specifically, we conclude that, for most possible ways we trim the catalogue, the results favour a bimodal shape of the posterior PDFs, where it can either be a truncated mass function with a break mass $M_\mathrm{break} \sim 10^{4.5}$ M$_\odot$ and a slope close to $-2$, experiencing early and mild disruption at $\sim 10^{6.5}$ yr, or a pure power law with a slope of $\sim -2.2$, with either no or very late disruption. Examining \autoref{fig:trimmed_mass} and \autoref{fig:trimmed_age}, it is clear that the 95\% confidence regions fur the full catalogue (black) in overlap with the 95\% confidence intervals of all the trimmed catalogues, and that the 95\% confidence intervals of all trimmed catalogues overlap with each other. Thus there is no strong evidence for differences. The exclusion we consider that has the largest effect is if we remove OGCs; doing so shifts the posteriors in favour of the first of the two probability peaks and away from the second, but even in this case the 95\% confidence intervals overlap and thus there is no convincing evidence for a change.

\subsection{CMF with a high mass truncation}
\label{sec:CMF}


Our results clearly favour a CMF that at high masses declines more sharply than the $\alpha_M = -2$ powerlaw commonly-found to describe cluster mass functions; our fits are ambivalent about what functional form best describes this, with one possibility being a powerlaw with a steeper slope $\alpha_M \approx -2.2$ and no exponential truncation, and the other being a powerlaw with slope $\alpha_M \approx -2$ but an exponential truncation at $M_\mathrm{break} \approx 10^{4.5}$. Given this finding, it is interesting to compare our $M_\mathrm{break}$ to values reported in the literature. Previous studies of CMF in spiral galaxies such as M83 \citep{2015GEM83}, NGC 628 \citep{LEGUS2017}, and M51 \citep{2018Messa} all found that Schechter functions provided better fits to the CMF than pure powerlaws, but with exponential truncations in mass at $\approx 10^{5.6}$ M$_{\odot}$, which agrees with our best-fitting truncation mass within an order of magnitude. However, given the bimodal nature of our posterior PDF, we cannot rule out either the possibility of a pure power law shape of the CMF, or that there is a truncation at a break mass $\approx 0.5-1$ dex smaller than this value. 

\begin{figure}
    \centering
    \includegraphics[width=\columnwidth]{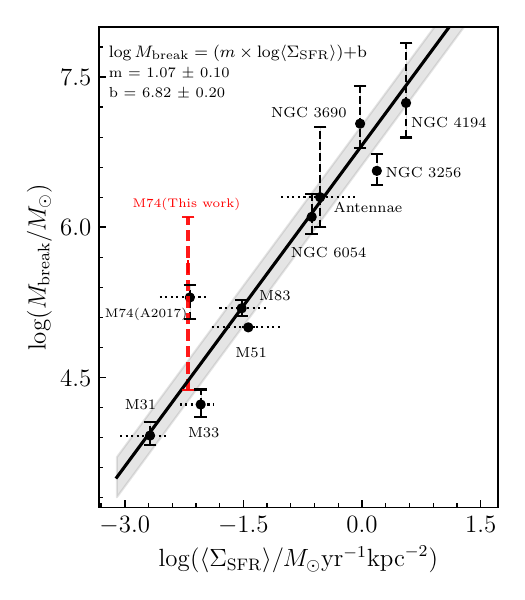}
    \caption{Comparison of break masses $M_\mathrm{break}$ and star formation rate surface densities $\Sigma_{\mathrm{SFR}}$ for a sample of nearby galaxies. Black points show the compilation of \citet{2022Wainer}, while the black line shows the best-fit linear relation obtained by \citet{2017PHAT}; the grey band shows the 84\% confidence interval around the relation. ``M74(A2017)'' denotes the truncated mass derived in \citet{LEGUS2017}. The red dashed line shows our fitted truncated mass for the full NGC 628 catalogue at the 84\% confidence interval. We find that the cluster mass function is well-described by either a Schechter function with a slope $\alpha_M \approx -2$ and a truncation at $M_\mathrm{break} \approx 10^{4.5}$ M$_\odot$ or a steeper slope $\alpha_M \approx -2.2$ but a substantially higher truncation mass; this uncertainty leads to the extended range of possible $M_\mathrm{break}$ values in our study. We take our value of $\Sigma_{\mathrm{SFR}}$ for NGC 628 from \citet{2023Sun}. }
    \label{fig:Mc_SFRD}
\end{figure}

It is also worth noting that we do not necessarily expect all galaxies to have the same truncation masses for their CMFs. \citet{2017PHAT} propose a relation between the break mass and the star formation rate density $\langle \Sigma_{\mathrm{SFR}} \rangle$, perhaps due to both varying systematically with mean interstellar pressure; \citet{2022Wainer} extend this proposal by adding new data. To place our fitted break mass into this context, we add our value for $M_\mathrm{break}$ in NGC 628 to \citeauthor{2022Wainer}'s compilation in \autoref{fig:Mc_SFRD}; for this purpose, we adopt a star formation rate density for NGC 628 from \citet{2023Sun}. As the figure shows, our best-fit break mass is consistent with the relation fit by \citeauthor{2017PHAT},
\begin{equation}
\log \frac{M_\mathrm{break}}{\mathrm{M}_\odot} = (1.07\pm 0.10)\log\frac{\langle\Sigma_\mathrm{SFR}\rangle}{\mathrm{M}_\odot\;\mathrm{yr}^{-1}\;\mathrm{kpc}^{-2}} + (6.82\pm 0.20),
\end{equation}
but given the extremely broad uncertainty implied by our bimodal posterior PDF, this is not a particularly strong statement. If the lower $M_\mathrm{break}$ turns out to be correct, the resulting value would agree well with the \citeauthor{2017PHAT} relationship, which in turn would provide a natural explanation for the small truncation mass: NGC 628 has a lower star formation rate per unit area, presumably indicative of a lower interstellar pressure. Notably, \citet{2022Wainer} find a significant underestimation of the upward bias in the truncated mass of NGC~628 fitted by \citet{LEGUS2017}. \citeauthor{2022Wainer}'s revised truncation mass associated with a larger error bar of approximately 1.1 dex, aligns more closely with our fitted truncated mass and wide range of uncertainties.

\subsection{Implications for cluster disruption}
\label{ssec:cluster_disruption}

We find that mass-independent disruption (MID) is a better description of the cluster age distribution than mass-dependent disruption (MDD) in all regions except for clusters with galactocentric radii $< R_{50}$, and that disruption is weak at young ages -- either no disruption at all occurs for $\approx 200$ Myr and only then becomes rapid ($\alpha_T \approx -1.4$), or disruption starts from young ages but is very mild ($\alpha_T \approx -0.3$. As with our results for the CMF, it is interesting to put this result into the context of previous studies. As summarised by \citet{2019ARAA}, the results of these studies have tended to depend strongly on the method of cluster catalogue construction, with authors who use ``exclusive'' catalogues that exclude objects that do not have compact, round morphologies favouring MDD and weak disruption, while those who use ``inclusive'' catalogues without such morphological filters favouring MID and more rapid cluster disruption, $\alpha_T \approx -1$. This difference arises because the morphological filters tend to preferentially affect the youngest clusters, so applying them flattens the age distribution. There is a great deal of debate in the literature, which we shall not revisit here, about the extent to which morphological filters artificially remove real star clusters from the sample, versus the extent to which a failure to filter leads to catalogues including large numbers of unbound associations that were never bound to begin with.

Because we include LEGUS class 3 sources in our catalogue, and thus are not applying morphological filters, one would naively have expected our results to match more closely those coming from the inclusive catalogues. However, our actual result -- MID but weak disruption, at least at ages below a few hundred Myr -- sits partway between the inclusive and exclusive
results, and this is likely at least in part because our method allows us to investigate significantly older ages than in previous studies. This makes us much less sensitive to the young ages where the uncertainty about morphological filtering is most acute. Our statistical signal is not dominated by the small number of clusters with ages $\lesssim 100$ Myr, but by the much larger number of clusters at somewhat older ages, where identification of clusters much less ambiguous since at these ages all bound stellar systems are relaxed. 

It is also interesting to compare our results to those for NGC 628 specifically, for which \citet{LEGUS2017} found mass-independent disruption of clusters in the inner pointing NGC 628c, but mass-dependent disruption at the outskirts in pointing NGC 628e. Since we  choose to analyse the cluster population as a whole or divide it by galactocentric radius, rather than by pointing, the samples we analyse are not completely identical. However, since the clusters belonging to NGC 628c (the inner region) constitute the great majority of the population, our model selection result for NGC 628 as a whole is consistent with the findings of \citeauthor{LEGUS2017}. There is possibly some tension in the fact that we favour MID even when we divide our catalogues by galactocentric radius, while \citeauthor{LEGUS2017} find MDD in NGC 628e, but since the divisions are not exactly the same we cannot make this statement more quantitatively. However, it does reinforce the point that our method has the advantage that, since we are forward modelling the completeness in each field, we can easily break up the clusters by any criterion of our choice to derive the demographics in different parts of the galaxy.

\section{Summary And Conclusion}
We use a novel method described by \cite{SLUG2019} to derive the demographics of the cluster population of the nearby spiral galaxy NGC 628 from unresolved photometry measured by the LEGUS survey \citep{CalzettiLEGUS2015}. This method has the advantage of integrating diverse observational data imaged using different filters or depths, without the need for binning or imposing extreme completeness cuts. As a result, our method allows us to analyze the roughly 1200 clusters in NGC 628 catalogued by the LEGUS survey, as opposed to the past work that was restricted to include only the most massive and youngest $\sim 30\%$ of the population. Our method also obviates the need to assign a unique mass and age to every individual cluster of the population, and therefore avoids the inevitable information loss incurred by reducing complex, sometimes multiply-peaked posterior probability distributions for individual cluster masses and ages to single values or even Gaussians. We validate the method by comparing the photometric distributions that emerge from our fits to the observed ones and finding good agreement, indicating that our approach can identify models for the cluster population that reproduce observations.
Our analysis results in three primary findings. 

First, we use the Akaike information criterion to compare two classes of parametric models for the time-evolution of the cluster population, mass-independent disruption (MID) and mass-dependent disruption (MDD). We find that the MID model fits the data better, both for the entire NGC 628 cluster catalogue and for all sub-divisions of it except one; the exception is if we consider only the half of the catalogue closer to the galactic centre. We thus conclude that cluster mass and age distributions are likely at least approximately separable and that the disruption time scales of the clusters have no dependence on the cluster masses. 

Second, we find that cluster mass functions are well-described by Schechter functions $dN/dM \propto M^{\alpha_{M}} \exp(-M/M_{\mathrm{break}})$, and that there is good evidence that the mass function at $\sim$$10^5$~M$_\odot$ is steeper than the slope $\alpha_M = -2$, most commonly found in previous studies. However, we are at present unable to determine if the data are better fit by an untruncated powerlaw with a steeper slope of $\alpha_M \approx -2.2$, or by a shallower slope but a truncation mass $M_\mathrm{break} \approx 10^{4.5}$~M$_\odot$. The latter possibility is consistent with the relationship between CMF truncation mass and star formation surface density proposed by \citet{2017PHAT} and \citet{2022Wainer}. 

Third, we find that the cluster age distribution in NGC 628 indicates that cluster disruption is weak at ages up to $\approx 200$ Myr; it either only begins at such ages and then becomes rapid, or begins early but then is very mild, with only a factor of $\approx 2$ decrease in cluster number per decade in age (index $\alpha_T \approx -0.3$). In past studies, the value of $\alpha_T$ has proven highly sensitive to the cluster catalogue type used, with studies using an inclusive cluster catalogues (those that do not include filters based on morphology) generally finding $\alpha_T \sim -1$ while those using exclusive catalogues (which are filtered for morphology) finding $\alpha_T \sim -0.3$. The significance of our study is that even though we are using an inclusive catalogue, we still find relatively weak disruption at young ages. One possible reason is that we can use a much larger sample and probe to considerably larger ages than previous studies such as \citet{LEGUS2017}. 

The three conclusions above are derived from our analysis of all the clusters in NGC 628 catalogued by LEGUS. However, we also group clusters by galactocentric radius and search for differences in cluster demographics with radius. We find marginal evidence for radial variations, with suggestive hints that inner galaxy clusters are subject to more severe disruption than those further out, and that inner galaxy clusters have a mass function that is better described by a pure powerlaw while outer galaxy ones have a mass function that is more strongly truncated. However, our ability to search for radial variations is limited by the fairly small number of clusters available at larger galactocentric radii, since the great majority of the available sample comes from the LEGUS NGC~628c pointing, which targets the galactic centre. Given the suggestive hints in our analysis, increasing the sample of outer-galaxy clusters would be a worthwhile effort.

Looking forward, now that we have demonstrated the performance of our pipeline on NGC 628, we will apply it to other cluster catalogues, both those originating from LEGUS and from other studies. Analysis of a diverse sample of galaxies will allow us to compare the demographics of cluster populations in various galactic environments. This in turn opens up the possibility of testing a range of theoretical models for cluster formation and disruption and their variation with the galactic environment.
\label{cha:conclusions}

\section*{Acknowledgements}

JT and KG are supported by the Australian Research Council through the Discovery Early Career Researcher Award (DECRA) Fellowship DE220100766 funded by the Australian Government. 
MRK acknowledges support from the ARC Future Fellowship and Laureate Fellowship funding schemes, awards FT180100375 and FL220100020.
This work is supported by the Australian Research Council Centre of Excellence for All Sky Astrophysics in 3 Dimensions (ASTRO~3D), through project number CE170100013. 
Based on observations made with the NASA/ESA Hubble Space Telescope, obtained at the Space Telescope Science Institute, which is operated by the Association of Universities for Research in Astronomy, Inc., under NASA contract NAS 5-26555. These observations are associated with program \#13364.

\section*{Data Availability}

The cluster catalog data underlying this article are publicly available online at
\href{https://archive.stsci.edu/prepds/legus/dataproducts-public.html}{MAST}. The software suite, tools, and LEGUS catalogues required to perform the MCMC optimization, as well as the output chains, are publicly accessible on \href{https://bitbucket.org/janet_jianling_tang/legus_slug23/src/master/}{Bitbucket}.



\bibliographystyle{mnras}
\bibliography{mnras} 



\appendix

\section{The cluster library}
\label{app:library}

The \citet{SLUG2019} approach described in the main text requires a library of synthetic clusters. We construct a library of $4 \times 10^7$ clusters for this purpose using the \textsc{slug} stochastic stellar population synthesis code \citep{2012SLUG, 2015SLUG}. Construction of the library involves two steps. In the first, we choose a mass $M$, age $T$, and extinction $A_\mathrm{V}$ for each cluster from a specified library distribution $p_\mathrm{lib}(M, T, A_\mathrm{V})$. We choose this library distribution to be
\begin{equation}
p_{\mathrm{lib}}\left(M,T,A_\mathrm{V}\right) \propto p(M) p(T) p\left(A_{\mathrm{V}}\right),
\end{equation}
with 
\begin{eqnarray}
p(M) & \propto &
\begin{cases} 
(M/10^2\,\mathrm{M}_\odot)^{-1}, \quad &
10^{2} < M/\mathrm{M}_{\odot} \leq 10^{5} \\
10^{-3} (M/10^5\,\mathrm{M}_\odot)^{-2}, & 10^{5}< M/\mathrm{M}_{\odot} \leq 10^{7} \\
0, & \text{otherwise}
\end{cases} 
\nonumber
\\
p(T) & \propto & 
\begin{cases}
T^{-1}, \quad & 10^{5} < T/\mbox{yr} < 1.5 \times 10^{10}\\
0, & \text{otherwise}
\end{cases}
\nonumber
\\
p\left(A_{\mathrm{V}}\right) &\propto & 
\begin{cases}
\text {const}, \quad &0<A_{\mathrm{V}}<3\text{ mag}\\
0, & \text{otherwise}
\end{cases}
\nonumber
\end{eqnarray}
This distribution maximises sampling of clusters with low masses and young ages, where stochastic effects are strongest and thus the largest number of samples is needed to fully characterise the luminosity distribution.

The second step is to generate photometric magnitudes for each cluster. To do so, we run \textsc{slug} using the Padova tracks with TP-AGB stars
\citep{2006Padova}, coupled to the ``starburst99'' treatment of stellar atmospheres \citep{Leitherer99a}. We draw stars from a Chabrier IMF using \textsc{slug}'s ``stop nearest'' sampling policy. Full explanations of these choices are provided in \citet{2015SLUG}, and in the \textsc{slug} documentation available at \url{https://slug2.readthedocs.io/}. The \textsc{slug} parameter file used to generate the library is available at \url{https://bitbucket.org/janet_jianling_tang/legus_slug23/src/master/}. This file includes all the necessary inputs and parameters described above. The library itself is too large to post, but is available upon request to the authors.

\section{The effects of stellar tracks}
\label{app:stellar_tracks_changes}

\begin{figure}
    \centering
    \includegraphics[width=0.99\columnwidth]{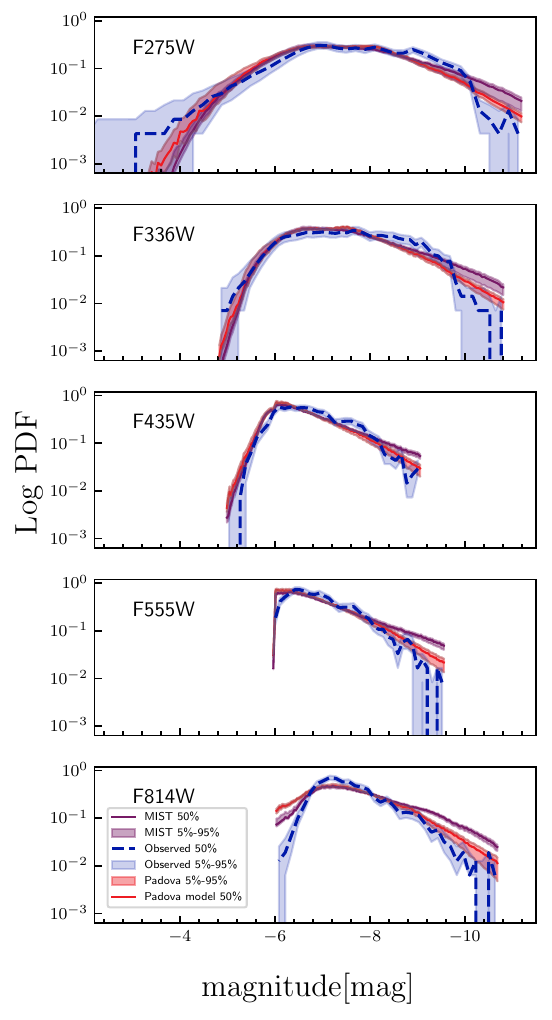}
    \caption{Same as \autoref{fig:1D_MID}, but now also showing luminosity functions generated using MIST stellar tracks (purple). The observed (blue) and Padova track fit (red) curves are identical to those shown in \autoref{fig:1D_MID}.}
    \label{fig:compare_models}
\end{figure}

To investigate how the choice of stellar tracks affects our analysis, we repeat our fit to the full catalogue, which in the main text we perform using a library generated from Padova-AGB stellar tracks \citep{2005PadovaAGB}, with a library generated from MIST stellar tracks \citep{2016MIST}. We leave all other aspects of the procedure unchanged. We then use the best-fitting results to generate 1D luminosity functions for the MIST models via the procedure described in \autoref{ssec:lum_func}, which we compare to the fiducial Padova results and to the observations in \autoref{fig:compare_models}. In this figure Padova-AGB and observed curves shown here are identical to those shown in \autoref{fig:1D_MID}, so the only difference is the addition of the MIST model results. Examining the figure, it is evident that the best-fitting Padova models reproduce the observed luminosity functions substantially better than the best-fitting MIST models, especially at high luminosity and in the redder V and I bands. It is this difference the leads to us adopting the Padova-AGB models as our preferred library for the analysis in the main text.

If we compare the best-fitting parameter values themselves, we find that, similar to the Padova fits, the fits using MIST tracks also prefer the MID over the MDD model. However, for MIST we find a significantly lower truncation mass, $M_{\mathrm{break}} \sim 10^4 M_{\odot}$, and a shallower mass slope of $\alpha_M \sim -1$. These changes combine so that the Padova and MIST models actually yield relatively little difference in the slope of the mass function near $M \sim 10^4$ M$_\odot$, where the LEGUS catalogue is most complete and sensitive -- that is, if we evaluate the slope $d\log p_M/d\log M$ at $M = 10^4$ M$_\odot$, then we obtain similar values close to $-2$ for both $\alpha_M \approx -2.2$, $M_\mathrm{break} \approx 10^{5.6}$ M$_\odot$, the $50^{\mathrm{th}}$ percentile values we find for the Padova models, and for the $\alpha_M \sim -1$, $M_{\mathrm{break}} \sim 10^4 M_{\odot}$ we obtain from MIST. The fits differ mainly away from this mass, where sensitivity is limited by small number statistics at the massive end and catalogue completeness at the low-mass end. The MIST result suggests a mass function that becomes relatively flat for masses $\ll 10^{4} M_{\odot}$ compared to the Padova CMF, but that drops off sharply at $\gg 10^4 M_{\odot}$ due to its lower break mass.

\section{Distribution of extinctions}
\label{app:Av}
\begin{figure}
    \centering
    \includegraphics[width=\columnwidth]{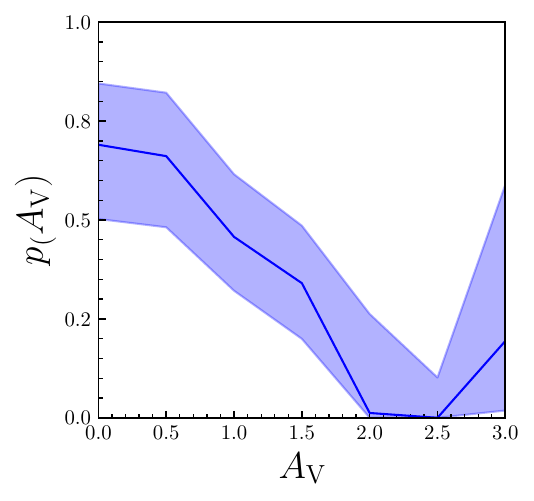}
        \caption{Posterior PDF of the distribution of dust extinctions $A_{\mathrm{V}}$. The solid line shows the $50^{\mathrm{th}}$ percentile result, and the shaded band shows the $5^{\mathrm{th}}$ to $95^{\mathrm{th}}$ percentile range.}
    \label{fig:AVPDF}
\end{figure}

As discussed in the main text, in addition to posteriors for our parameters of interest, our fit also necessarily produces posteriors for the distribution of dust extinctions. While this is a nuisance parameter, it is important for us to verify that the results is reasonable. We show the 1D marginal posterior PDF of dust extinctions that we derive for our MID model fit to the full LEGUS catalogue in \autoref{fig:AVPDF}. We see that the results are in reasonable agreement with expectations: the extinction is small to moderate for most clusters, $A_V \lesssim 1$, with a minority having higher extinction. We have confirmed that the upturn in $A_V$ is not due to outliers in photometric space.

We note that uncertainty about the shape of the extinction distribution does contribute to the overall uncertainty of our fits, since the posterior PDFs for all quantities that we care about must be marginalised over the nuisance $A_V$ distribution. In particular, as discussed in \citet{2015SLUG}, star clusters can exhibit degeneracies between mass, age, and extinction, leading to multiple fits of comparable likelihood with varying sets of physical parameters. In such cases, utilising a full PDF as done in this work, provides more realistic estimate of the errors as compared to traditional SED fitting methods that return a single best fit with a Gaussian error estimate. However, there is certainly still room for improvement, In particular, \citet{2017MNRAS.469.2464A} has demonstrated that the inclusion of further photometric data, particularly narrow-band H$\alpha$, can help break some of these degeneracies in forward-modelling techniques. Presumably the addition of such data to our population-level fits would also help reduce the uncertainties.

\section{ADDITIONAL COMPARISONS IN PHOTOMETRIC SPACE}
\label{app:MDDcomp}

\begin{figure}
    \centering
    \includegraphics[width=\columnwidth]{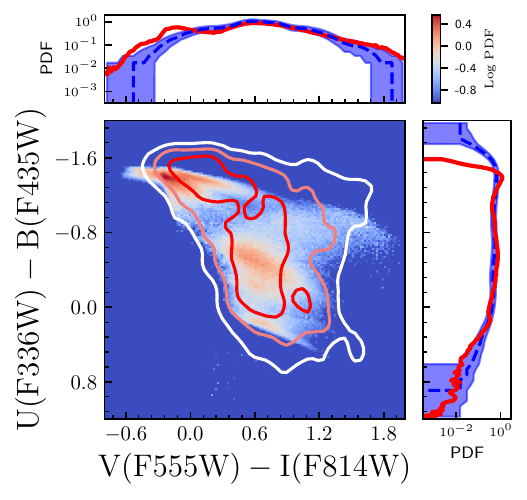}
    \caption{Same as \autoref{fig:UV_UVU}, but showing the U - B versus V - I colour-colour distributions of the observations and the best-fitting model. The y-axis is inverted to align with conventions used in star cluster evolutionary tracks, such as Figure 20 of \citet{2022Deger}
    .}
    \label{fig:UB_VI}
\end{figure}

\begin{figure}
    \includegraphics[width=\columnwidth]{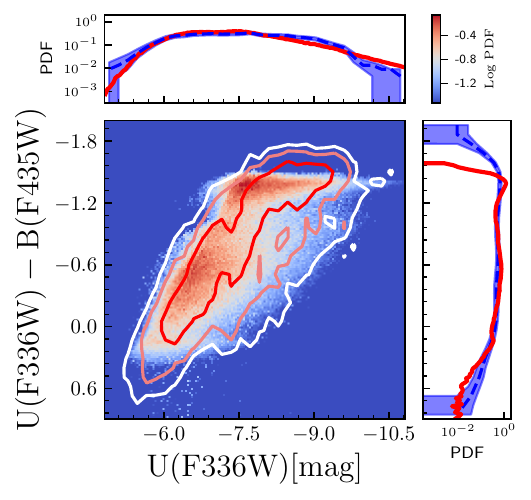}
    \caption{Same as \autoref{fig:UV_UVU}, except we show U versus U-B in colour-magnitude space here.}
    \label{fig:U_UB}
\end{figure}

\begin{figure}
    \centering
    \includegraphics[width=\columnwidth]{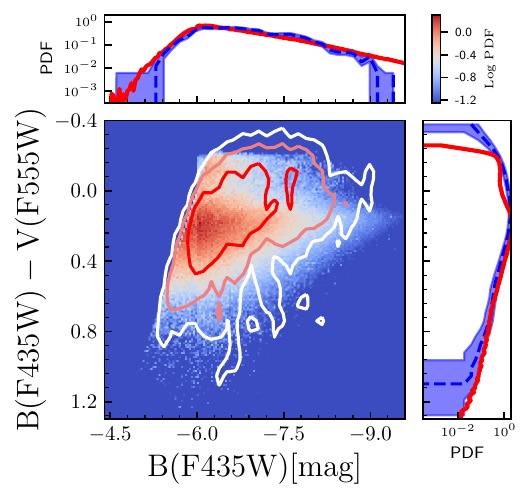}
    \caption{Same as \autoref{fig:UV_UVU}, except we show B versus B-V in colour-magnitude space here.}
    \label{fig:B_BV}
\end{figure}

\begin{figure}
    \includegraphics[width=\columnwidth]{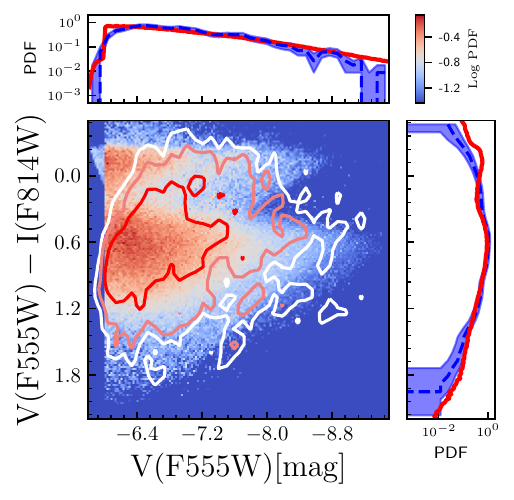}
    \caption{Same as \autoref{fig:UV_UVU}, except we show V versus V-I in colour-magnitude space here.}
    \label{fig:V_VI}
\end{figure}

In addition to the colour-colour and colour-magnitude plots of UV-U versus U-B and UV-U versus U presented in Section \autoref{ssec:photocomparison}, where the most significant disagreements occur, we also present colour-magnitude comparisons for all the other bands in \autoref{fig:U_UB}, \autoref{fig:B_BV}, and \autoref{fig:V_VI}.

\bsp	
\label{lastpage}
\end{document}